\documentclass[acs,jctc,superscriptaddress,amsmath,amssymb,twocolumn]{revtex4-2}

\usepackage[utf8]{inputenc}
\usepackage{amsmath}
\usepackage{comment}
\usepackage{mathtools}
\usepackage{xcolor}
\usepackage[unicode]{hyperref}
\usepackage{revsymb}
\usepackage{tikz}
\usepackage{graphicx}
\usepackage{siunitx}
\usepackage{multirow}
\usepackage[dvipsnames]{xcolor}
\usetikzlibrary{arrows, arrows.meta}
\usetikzlibrary{math} 
\usetikzlibrary{decorations.markings}
\usetikzlibrary { decorations.pathmorphing, decorations.pathreplacing, decorations.shapes, }
\tikzset{snake it/.style={decorate, decoration=snake}}
\usepackage{dcolumn}
\newcolumntype{d}[1]{D{.}{.}{#1}}

\hypersetup{
   unicode=true,          
   plainpages=false,
   colorlinks=true,       
   linkcolor=blue,          
   citecolor=blue,        
}
\usepackage{url}

\newcommand{\bos}[1]{\boldsymbol{#1}}

\def\Eh{E_\text{h}}
\def\cm{\text{cm}^{-1}}

\def\br{\boldsymbol{r}}
\def\iim{\text{i}}
\def\tT{\text{T}}

\def\SM{\text{SI}}

\def\nel{n_\text{el}}

\def\atSup{$\text{a}\ ^3\Sigma_\text{u}^+$}

\def\XdSup{$\text{X}\ ^2\Sigma_\text{u}^+$}
\def\XdSgp{$\text{X}\ ^2\Sigma_\text{g}^+$}
\def\btPg{$\text{b}\ ^3\Pi_\text{g}$}
\def\ctSgp{$\text{c}\ ^3\Sigma_\text{g}^+$}

\def\Eh{E_\text{h}}
\def\cm{\text{cm}^{-1}}

\def\br{\boldsymbol{r}}
\def\iim{\text{i}}
\def\Nb{N_\text{b}}

\def\nnuc{N_\text{nuc}}

\def\MV{\text{MV}}
\def\Done{\text{D1}}
\def\Dtwo{\text{D2}}
\def\OO{\text{OO}}
\def\bp{\boldsymbol{p}}
\def\bs{\boldsymbol{s}} 

\def\hbp{\hat{\boldsymbol{p}}}
\def\hbs{\hat{\boldsymbol{s}}} 

\def\br{\boldsymbol{r}}

\def\bA{\boldsymbol{A}}
\def\hH{\hat{H}}

\def\vphi{\varphi}
\def\brho{\bos{\rho}}
\def\el{\text{el}}
\newcommand{\pd}[2]{\frac{\partial #1}{\partial #2}}

\definecolor{ao}{rgb}{0.0, 0.5, 0.0}

\usepackage{setspace}

\usepackage[unicode]{hyperref}
\usepackage{soul}
\hypersetup{
   unicode=true,          
   plainpages=false,
   colorlinks=true,       
   linkcolor=black,          %
   linkcolor=blue,          
   citecolor=blue,        
   urlcolor=blue          
}

\begin{document}

\title{%
Rovibrational computations for the He$_2$ \atSup\ state including non-adiabatic, relativistic, and QED corrections
}

\author{Ádám Margócsy}
\affiliation{MTA–ELTE Lendület `Momentum' Molecular Quantum electro-Dynamics Research Group,
Institute of Chemistry, Eötvös Loránd University, Pázmány Péter sétány 1/A, Budapest, H-1117, Hungary}

\author{Balázs Rácsai}
\affiliation{MTA–ELTE Lendület `Momentum' Molecular Quantum electro-Dynamics Research Group,
Institute of Chemistry, Eötvös Loránd University, Pázmány Péter sétány 1/A, Budapest, H-1117, Hungary}

\author{Péter Jeszenszki}
\affiliation{MTA–ELTE Lendület `Momentum' Molecular Quantum electro-Dynamics Research Group,
Institute of Chemistry, Eötvös Loránd University, Pázmány Péter sétány 1/A, Budapest, H-1117, Hungary}

\author{Edit Mátyus}
\email{edit.matyus@ttk.elte.hu}
\affiliation{MTA–ELTE Lendület `Momentum' Molecular Quantum electro-Dynamics Research Group,
Institute of Chemistry, Eötvös Loránd University, Pázmány Péter sétány 1/A, Budapest, H-1117, Hungary}

\date{\today}

\begin{abstract}
\noindent
A potential energy curve (PEC) accurate to a fraction of 1 ppm ($1:10^6$) is computed for the \atSup\ state of He$_2$ endowed with relativistic and QED corrections. The nuclear Schrödinger equation is solved on this PEC with diagonal Born--Oppenheimer and non-adiabatic mass corrections to obtain highly accurate rotational-vibrational levels. 
The computed rovibrational intervals and fine-structure splittings, spanning over several orders of magnitude in energy, are found to be in remarkable agreement with available high-resolution spectroscopy data. 
\end{abstract}

\maketitle

\section{Introduction}
 \noindent %
Spectroscopic studies on few-particle atoms and molecules often serve as important benchmarks and precision tests of current physical theories, as well as a possible way to refine the values of physical constants. 

Spectroscopists have been interested in the low-lying electronically excited (Rydberg) states of $\text{He}_2$ for decades. Compared to the very weakly bound $\text{X} \ ^1\Sigma_\text{g}^+$ ground state of He$_2$ (supporting a single bound rovibrational state~\cite{RySzJe89,TaToYi95,GrScToHeKoSt00,CePrKoMeJeSz12,SuVaBook98,SzJe10,MaRe12,MiBuHoSuAdCeSzKoBlVa13}), the excited states experience a much stronger binding, leading to rich rovibrational and magnetic properties. The triplet He$_2$ states are relatively long-lived due to their dominant decay channels to the very weakly bound singlet X state being spin-orbit coupling mediated radiative transitions \cite{ChJeYaLe89,McBrBuDzHuMaDoGoHa99}. 

The accuracy of experiments has improved significantly over the years for this few-electron homonuclear diatomic molecule, the uncertainties of the measured ionization energy \cite{SeJaCaMeScMe20}, vibrational spacings~\cite{BrGi71,GiGi80,RaScVaMe08,HoWiShBeMe25}, rotational intervals~\cite{SeJaMe16,SeJaCaMeScMe20,SemeriaPhD2020}, fine-structure splittings of rovibrational levels~\cite{LiMcVi74,FoBeCo98,RoBrBeBr88,SeJaClAgScMe18,WiHoMe25} have been reduced from $\sim10^{-3}$~cm$^{-1}$ down to $10^{-9}$ \, \text{cm}$^{-1}$, reaching the sub-kHz regime in extreme cases. With new techniques like laser cooling on the horizon \cite{VeZeKnRoBe25}, further high-quality experimental data can be foreseen.

At the same time, theoretical predictions lag (much) behind in many aspects. Probably the first PEC computation for the $\text{He}_2 \ \text{`a'}$ state was done by Buckingham and Dalgarno in 1952 \cite{BuDa52}, which was superseded only almost ten years later \cite{BrBrMa61}. Even nowadays, mainly older results are available for the {\atSup} state obtained with orbital-based quantum chemical (\emph{e.g.,} MC-SCF) approaches
~\cite{BeNiMu74,Co76,SuLiSiSiJoSh83,KoLe87,ChJeYaLe89,Ya89},
 which show a non-negligible discrepancy with experiment;  the more recent
techniques~\cite{McGiBuDa93,NiKrPrViWi2019,EpMoChTe24,XuLuZhGuSh24},
while often interesting on their own right, cannot deliver the required accuracy either.
One of the main bottlenecks is the use of (standard) uncorrelated atomic basis sets: even quite large augmented correlation-consistent basis sets can barely bring the error of the non-relativistic electronic energy below $\sim1 \, \text{m}\Eh$, raising the need for an explicitly correlated \cite{Bo60,SzAdSa79,AlMoSz86,AlMoSz87,AlMoSz88,CeRy93,CeRy95} PEC computation (although this view may be challenged in the future \cite{DaGrPrTo25}). To the best of our knowledge, only a single-point explicitly correlated computation has been performed for the \atSup\ state so far~\cite{PaCaBuAd08}.
Furthermore, relativistic, quantum electrodynamical (QED) and non-adiabatic effects also must be taken into account to match with experiments.
The present work significantly improves upon previous computational results and finally reaches a close agreement with the available experimental data. 

\section{The rovibronic problem}
The  Schrödinger equation describing the internal motion of a homonuclear diatomic molecule with nuclear charge $Z$, nuclear mass $M_\text{nuc}$, and $\nel$ electrons reads
\begin{align}
  \left[-\frac{1}{2\mu} \boldsymbol{\nabla}^2_{\bos{\rho}}
    + \hat{H}_\el(\brho)
    +\frac{1}{8\mu} \hat{\boldsymbol{P}}^2_\el\right] \Psi(\br,\brho) &= E \Psi(\br,\brho) \ ,
\end{align}
where $\hat{\boldsymbol{P}}_\el=\sum_{i=1}^{n_\el} \hat{\bp}_i=-\iim\sum_{i=1}^{n_\el}\boldsymbol{\nabla}_i$, $\brho=\boldsymbol{R}_{1}-\boldsymbol{R}_{2}$, $(\boldsymbol{\nabla}_{\bos{\rho}})_i=\partial/\partial{\rho_i}$ and  $\mu=M_\text{nuc}/2$. Obtaining the accurate eigenvalues of this equation
goes in three steps in the Born--Oppenheimer (BO) framework: 
(1) solving the electronic eigenvalue problem of
\begin{align}
  \hat{H}_{\text{el}}(\brho)
  =& 
  -\frac{1}{2} 
  \sum_{i=1}^{\nel} 
  \boldsymbol{\nabla}_i^2
  -\sum_{i=1}^{\nel}\left[\frac{Z}{\left|\br_i-\frac{1}{2}\brho\right|}
  +\frac{Z}{\left|\br_i+\frac{1}{2}\brho\right|}\right] \nonumber \\
  &+ 
\sum_{i=1}^{\nel}\sum_{j=i+1}^{\nel}
  \frac{1}{\left|\br_i-\br_j\right|} 
  +\frac{Z^2}{\rho}
  \ ;
  \label{Hel}
\end{align}
(2) improving the electronic energy with adiabatic nuclear mass, relativistic and QED corrections;
(3) solving for nuclear (rovibrational) degrees of freedom using the corrected PEC and nonadiabatic mass corrections. 
These steps are presented below with numerical results.

\section{Solving the electronic problem \label{sec:sch_ele}}

The electronic wave function is the solution of
\begin{equation}
 \hat{H}_{\text{el}}(\brho)\vphi_\text{a}^{(\Sigma)}(\br;\brho)=U_\text{a}(\rho)\vphi_\text{a}^{(\Sigma)}(\br;\brho) \ ,
\end{equation}
the $\Sigma$ spin projection being along the internuclear axis $\boldsymbol{e}_z=\brho/\rho$ (not to be confused with the spatial symmetry label $\Sigma_\text{g/u}$), and index `a' referring to \atSup. We parametrize $\vphi_\text{a}^{(\Sigma)}(\br;\brho)$ as a truncated expansion of floating explicitly correlated Gaussians~(fECG-s)~\cite{SuVaBook98,SzJe10,MaRe12,MiBuHoSuAdCeSzKoBlVa13}, turning the electronic eigenvalue problem into a generalized matrix eigenvalue problem.
The $U_\text{a}(\rho)$ potential energy curve (PEC) was computed at several points over the range $\rho/a_0\in[1,100]$ starting from a thoroughly optimized point at $2\, a_0$, and always using the (rescaled) wave function parameters of the previous point as an initial guess for the next one \cite{CeRy95,FeMa19HH,FeKoMa20,FeMa22h3}. The curve was computed with step size $0.05 \, a_0$ for $\rho/a_0\in[1,10]$, $0.1 \, a_0$ for $\rho/a_0\in[10,16.5]$ and $1\, a_0$ for $\rho/a_0\in[17,100]$. 
The computed (variational) electronic energy is an upper bound to the exact value; the convergence error for the $N_\text{b}=1500$ basis set (used for the PEC generation) is estimated to be $\sim 3 \, \mu\Eh$ at $\rho=2 \, a_0$ (near the equilibrium structure), while it is smaller, $\sim 0.2 \, \mu\Eh$ at $\rho=100 \, a_0$ (the large $\rho$ limit can be compared with the known atomic results \cite{AzBeKo18}). Further computations at $\rho=2 \, a_0$, including up to 2500~fECGs, reduce the convergence error to $\sim1 \, \mu\Eh$.
The Supporting Information (\SM{}) reports further computational details, convergence studies and a comparison with standard quantum chemistry methods. 

\section{Correcting the electronic energy \label{sec:ele_corr}}
The corrected PEC is written as
\begin{equation}
 W=U+U_{\text{DBOC}}
 +U_{\text{rel+QED,sn}}+U_{\text{f.nuc.}} \ ,
\end{equation}
\emph{i.e.,} $U(\rho)$ endowed with diagonal Born--Oppenheimer, spin-independent (sn) relativistic \& QED and finite nuclear size corrections \emph{(vide infra)}. There is a further important, spin-dependent (sd) relativistic \& QED contribution (denoted by $U^{(\Sigma)}_{\text{sd}}$), which is treated independently and not defined to be part of the PEC.

\subsection{Finite nuclear mass corrections}
 
The diagonal Born--Oppenheimer correction (DBOC) consists of electronic, vibrational and angular momentum parts: 
\begin{equation}
    U_{\text{DBOC}}=
    \frac{1}{8\mu} 
    \langle \hat{\boldsymbol{P}}^2_\el \rangle
    -\frac{1}{2\mu}
    \left\langle \frac{\partial^2}{\partial \rho^2} \right\rangle
    +
    \frac{1}{2\mu\rho^2}\langle \hat{L}^2_x + \hat{L}^2_y \rangle \ ,
\end{equation}
where $\hat{\boldsymbol{L}}$ is the electronic angular momentum; convergence tests are reported in the \SM{}.
We note that this internal-coordinate form is identical to the more common Cartesian form of the DBOC, as it was shown by Cencek and Kutzelnigg~~\cite{Ku97,CeKu97}. We use here the internal-coordinate form to be able to exploit the point-group symmetry with the fECG basis set.

It is appropriate to introduce here the vibrational and rotational non-adiabatic mass corrections~\cite{Ma18nonad,MaTe19}, although treated as a correction to the kinetic energy, not the PEC:
\begin{align}
  \delta m^\text{vib}
  =4
  \left\langle 
    \pd{\vphi}{\rho_z} \left|
      (\hat{H}_\el-U)^{-1} P^\perp
    \right| \pd{\vphi}{\rho_z}
  \right\rangle \ ,
\end{align}
\begin{align}
  \delta m^\text{rot}
  &=4
  \left\langle 
    \pd{\vphi}{\rho_\beta} \left|
      (\hat{H}_\el-U)^{-1} P^\perp
    \right| \pd{\vphi}{\rho_\beta}
  \right\rangle \ ,
\end{align}
with the projector $P^\perp = 1- | \vphi\rangle\langle \vphi|$, and $\beta=x,y$. Their role is to incorporate coupling with distant electronic states perturbatively, by formally modifying the reduced mass in the nuclear kinetic energy term (\emph{vide infra}),
\begin{align}
  \frac{1}{2\mu^\text{v/r}(\rho)}
  =
  \frac{1}{2\mu}
  \left[%
    1- \frac{\delta m^\text{vib/rot}(\rho)}{2\mu}
  \right] \ .
\end{align}

\subsection{Spin-independent relativistic \& QED and finite nuclear size corrections}
 
The correction $U_{\text{rel+QED,sn}}$ for a given electronic state is computed as
\begin{equation}
U_{\text{rel+QED,sn}}=U_{\text{rel}}+U_{\text{lQED}}+U_{\text{hQED}}
\ .
\label{Ecentroid}
\end{equation}
The leading relativistic correction $U_{\text{rel}}$ is obtained as the expectation value of the (spin-independent) Breit--Pauli Hamiltonian $\alpha^2\hH_{\text{BP,sn}}$, where
\begin{align}
  \hH_{\text{BP,sn}} 
  =
  \hH_\MV + \hH_\Done +  
  \hH_\OO + \hH_\Dtwo + \hH_\text{SS,c} 
  \label{eq:Hbp0}
\end{align}
contains the mass-velocity correction,
\begin{align}
  \hH_\MV
  =
  -\frac{1}{8} 
  \sum_{i=1}^{\nel} (\hbp_{i}^{2})^2 \ ,
\end{align}
the one-electron Darwin term,
\begin{align}
  \hH_\Done
  =
  \frac{\pi}{2}\hat{\delta}_1 \ \ , \ \ 
  \hat{\delta}_1=\sum_{i=1}^{\nel}
  \sum_{A=1}^{\nnuc}
    Z_A\delta(\br_{iA}) \ ,
\end{align}
the orbit-orbit interaction,
\begin{align}
  \hH_{\OO}
  =
  -\frac{1}{2}
  \sum_{i=1}^{\nel}\sum_{j=i+1}^{\nel}
    \left[%
      \frac{1}{r_{ij}}\hbp_{i}\hbp_{j} + \frac{1}{r_{ij}^{3}}(\br_{ij}(\br_{ij}\hbp_{j})\hbp_{i})
    \right] \ ,
\end{align}
the two-electron Darwin term,
\begin{align}
  \hH_{\Dtwo}
  =
  -\pi\hat{\delta}_2 \ \ , \ \ 
  \hat{\delta}_2=
  \sum_{i=1}^{\nel}\sum_{j=i+1}^{\nel}
    \delta(\br_{ij}) \ ,
\end{align}
and the Fermi contact part of the spin-spin interaction (with $\hat{\boldsymbol{s}}_i=I\otimes ...\otimes (\boldsymbol{\sigma}/2)\otimes...\otimes I$),
\begin{align}
  \hat{H}_\text{SS,c} 
  =  
  -\frac{8\pi}{3}\sum_{i=1}^{\nel}\sum_{j=i+1}^{\nel}
    \hbs_i \hbs_j \delta(\br_{ij}) \ .
\end{align}
We note that $\hat{H}_\text{SS,c}|\psi\rangle=2\pi \, \hat{\delta}_2|\psi\rangle$ for any antisymmetric wave function (see Ch. 2 of Ref. \cite{HeJoOlbook}), leading to 
\begin{align}
\hH_{\text{BP,sn}} 
  =&
  -\frac{1}{8} 
  \sum_{i=1}^{\nel} (\hbp_{i}^{2})^2 + \frac{\pi}{2}\hat{\delta}_1+\pi\hat{\delta}_2   \nonumber \\
  &-\frac{1}{2}
  \sum_{i=1}^{\nel}\sum_{j=i+1}^{\nel}
    \left[%
      \frac{1}{r_{ij}}\hbp_{i}\hbp_{j} + \frac{1}{r_{ij}^{3}}(\br_{ij}(\br_{ij}\hbp_{j})\hbp_{i})
    \right]  \ .
  \label{eq:Hbp} 
\end{align}
The $\alpha^3\Eh$ energy correction arises from non-radiative (virtual pair / photon exchange) and one-loop radiative QED effects (self-energy, vertex corrections, and vacuum polarization). The final energy correction $U_{\text{lQED}}$ is computed as the expectation value~\cite{JeAdBook22, sucherPhD1958, Ar57}, 
\begin{align}
  U_{\text{lQED}}
  =&\alpha^3
  \Bigg\langle \frac{4}{3}\left[\frac{19}{30}-2\ln(\alpha)-\ln(k_0) \right]\hat{\delta}_1 \nonumber \\
   &+ \left[\frac{164}{15}+\frac{14}{3}\ln(\alpha) \right]\hat{\delta}_2
    -\frac{7}{6\pi}
  {\cal{P}}\left(\frac{1}{r^3}\right)  \Bigg\rangle \ ,
  \label{E3}
\end{align}
where $\ln(k_0)$ is the (state-specific, non-relativistic) Bethe logarithm:
\begin{equation}
 \ln(k_0)=\frac{\left\langle\hat{\boldsymbol{P}}_\el(\hat{H}_\text{el}-U)\ln(2|\hat{H}_\text{el}-U|)\hat{\boldsymbol{P}}_\el\right\rangle}{2\pi\langle\hat{\delta}_1\rangle} \ ,
 \label{BLterm}
\end{equation}
and ${\cal{P}}(1/r^3)$ is the Araki--Sucher distribution (with the Euler-Mascheroni constant $\gamma$):
\begin{align}
 &{\cal{P}}\left(\frac{1}{r^3}\right)=
 \nonumber \\ 
 &\lim_{\epsilon\rightarrow0^+}\sum_{i=1}^{\nel}\sum_{j=i+1}^{\nel}\left[\frac{\Theta(r_{ij}-\epsilon)}{r_{ij}^3}+4\pi [\gamma+\ln(\epsilon)]\delta(\boldsymbol{r}_{ij})\right] \ .  
 \label{ASterm}
\end{align}
Higher-order QED effects are commonly estimated \cite{PuPa06,KoPiLaPrJePa11,PuKoPa15,PuKoCzPa16}
with the next-to-leading order part of the radiative one-loop correction 
from the hydrogenic theory \cite{KaKlSc52,BaBeFe53,Eides2001},
\begin{align}
  U_\text{hQED}=\pi\alpha^4\left[\frac{427}{96}-
  2\ln (2) \right]\sum_{A=1}^{\nnuc}Z^2_A\sum_{i=1}^{\nel}\langle\delta(\br_{iA})\rangle \ .
\end{align}
The finite nuclear size effect is calculated using \cite{Eides2001,JeAdBook22}
\begin{align}
  U_\text{f.nuc.}
  =
  \frac{2\pi}{3}\alpha^2\sum_{A=1}^{\nnuc}Z_A\frac{{\cal{R}}^2_A}{\lambdabar^2_\text{C}}\sum_{i=1}^{\nel}\langle\delta(\br_{iA})\rangle \ ,
  \label{eq:fnuc}
\end{align}
${\cal{R}}^2/\lambdabar^2_\text{C}$ being the mean squared radius of $^4$He$^{2+}$ in units of the reduced Compton wavelength.

It should be noted that on our current level of precision, neither the effect of $U_\text{hQED}$ nor that of $U_\text{f.nuc}$ is significant for the `a' state of He$_2$; the only reason we include them in the present computations is to be consistent with the more accurate He$_2^+$ computations of Ref.~\cite{paper-he2p} when evaluating the ionization energy in Sec.~\ref{sec:rovib}.

\subsection{Spin-dependent relativistic  \&  QED corrections}
 
The effect of electron spin-orbit, spin-own-orbit and spin-other-orbit interactions is zero for electronic states of $\Sigma_\text{g/u}$ symmetry; the only relevant spin-dependent effect in the present case is the magnetic dipole-dipole interaction: 
\begin{align}
 \alpha^2\hat{H}_\mathrm{sd}=\alpha^2\left(\frac{g}{2}\right)^2\hat{H}_\mathrm{SS,dp}\approx\alpha^2\left(1+\frac{\alpha}{\pi}\right)\hat{H}_\mathrm{SS,dp} \ ,
 \label{Hspindep}
\end{align}
where 
\begin{align}
  \hat{H}_\mathrm{SS,dp}
  =
  \sum_{i=1}^{\nel}\sum_{j=i+1}^{\nel}
    \left[%
      \frac{\hbs_{i} \hbs _{j}}{r_{ij}^3}
      -3\frac{\left(\hbs_{i}\bos{r}_{ij}\right)\left(\hbs_{j}\bos{r}_{ij}\right)}{r_{ij}^5}
  \right] \ .
\end{align}
QED manifests itself in modifying the free-electron $g$ factor,
\begin{equation}
 g=2\left[1+\frac{\alpha}{2\pi}+{\cal{O}}(\alpha^2)\right] \ .
\end{equation}
These $g$-factor corrections are automatically included in the form factor treatment of the Breit Hamiltonian ~\cite{JeAdBook22}. 

The $(2S+1)$-dimensional matrix representation of $\alpha^2\hat{H}_\mathrm{sd}$ over the $\varphi^{(\Sigma)}_\text{a}\ (\Sigma=-1,0,+1)$ subspace is diagonal (similarly for any  $\Sigma_\text{g/u}$ triplet states), 
since $\hat{H}_\mathrm{sd}$ commutes with $(\hat{\boldsymbol{L}}+\hat{\boldsymbol{S}})^2$ and $\hat{L}_z+\hat{S}_z$ (where $z$ is the body-fixed axis connecting the nuclei); the three {\atSup} states are eigenfunctions of $\hat{L}_z+\hat{S}_z$ with eigenvalue $\Sigma$. 
There is thus no coupling in this case, but an independent shift 
\begin{align}
  U^{(\Sigma)}_{\text{sd}}
  =
  \alpha^2\langle \varphi^{(\Sigma)}|\hat{H}_\text{ss,dp}|\varphi^{(\Sigma)}\rangle
\end{align}
for each state within the $(2S+1)$-manifold. It follows from angular momentum algebra that $\hat{H}_\mathrm{SS,dp}$ does not split the $\Sigma=\pm1$ states in this diatomic system; furthermore, the correction $U^{(\Sigma)}_{\text{sd}}$ does not shift the centroid of the energy levels, meaning (for triplet states) 
\begin{equation*} 
  U^{(\pm1)}_{\text{sd}}=-\frac{1}{2}U^{(0)}_{\text{sd}} \ .
\end{equation*}
The zero-field splitting of these states is defined as 
\begin{equation}
  \Delta E_\text{zfs}=U^{(0)}_{\text{sd}}-U^{(\pm1)}_{\text{sd}}=\frac{3}{2}U^{(0)}_{\text{sd}} \ .
  \label{zfseq}
\end{equation}
\newline
Through $\hat{H}_\text{sd}$, the coupling of the a~$^3\Sigma_\text{u}^+$ state with $^3\Pi_\text{u}$ and $^1\Pi_\text{u}$ states is possible. However, the closest f~$^3\Pi_\text{u}$ and F~$^1\Pi_\text{u}$ electronic states of He$_2$ are much higher in energy \cite{EpMoChTe24} ($>0.1$~$\Eh$), so explicit coupling was neglected in this work.
 
In the second part of the paper, the computed rovibrational fine-structure intervals will be compared with experimental data. 
In part, the comparison is made with reconstructed data through the effective Hamiltonian approach with parameters fitted to measured data (as outlined in the Appendix), but without including the coupling of electron spin and rotation magnetic moments. The combined effect of relativistic and nonadiabatic corrections (relativistic recoil) is also not considered, and is left for future work.

\subsection{Computations, numerical results}
 
Computing relativistic and QED corrections in the non-relativistic QED (nrQED) framework (using the non-relativistic state as reference state) involve the expectation values of highly singular operators (notably $\hat{\delta}_1$, $\hat{\delta}_2$, $\hat{H}_\text{MV}$ and ${\cal{P}}(1/r^3)$), which may lead to extremely slow convergence towards the complete basis limit when directly evaluated in a Gaussian basis. 
We note that an alternative theoretical framework, based on a correlated-relativistic reference state \cite{JeFeMa21,JeFeMa22,FeJeMa22b,JeMa23,MaFeJeMa23,MaMa24,NoMaMa24} instead of a non-relativistic reference is currently under development in our group, and it appears to be numerically more robust in finite basis computations. However, for the time being and for the sake of direct comparison with experimental data, all corrections are to be evaluated in the nrQED framework in which the Breit--Pauli Hamiltonian provides the leading-order relativistic corrections.

To compute expectation values of singular operators, instead of direct evaluation, we use regularization techniques. The so-called `Drachmanization'~\cite{Dr81,PaCeKo05}, and its numerically robust variation (numDr)~\cite{RaFeMaMa24,paper-he2p} was used to compute $\langle\hat{\delta}_1\rangle$, $\langle\hat{\delta}_2\rangle$, $\langle\hat{H}_\text{MV}\rangle$ and $\langle{\cal{P}}(1/r^3)\rangle$ along the PEC; expectation values of the other, less singular operators were computed in the standard way.
Further details are collected in the \SM{}, including convergence tables of specific values of $\langle\hat{\delta}_1\rangle$, $\langle\hat{\delta}_2\rangle$, $\langle\hat{H}_\text{MV}\rangle$ and $\langle{\cal{P}}(1/r^3)\rangle$ at the $2\, a_0$ point, and a comparison with the so-called Integral Transformation approach, another convergence acceleration technique \cite{PaCeKo05,JeIrFeMa22}.

The direct computation of the Bethe logarithm is also plagued by very slow basis set convergence, and the highly accurate approach of Schwartz~\cite{Sc61,KoHiKa13,Ko19,FeMa22bethe} is hindered by multiple costly basis set optimization steps. Numerical observations about the weak dependence of $\ln(k_0)$ on ${\nel}$ and the electronic quantum numbers motivated an ion-core approximation in Ref.~\cite{FeKoMa20}, where the Bethe logarithm of $\text{He}_2^+(^2\Sigma_\text{u}^+)$ was approximated with that of $\text{He}_2^{3+}(^2\Sigma_\text{g}^+)$ (a one-electron state). To avoid the costly evaluation of $\ln(k_0)$ using the Schwartz method, we use the same approximation for the \atSup\ state as well. Some reference values for atomic Bethe logarithms, as well as the ion-core and the rigorously calculated values for the \atSup\ diatomic state (at $\rho=2~a_0$) are displayed in the \SM{}. A comparison of the values in Table S5 suggests the ion-core approximation to be fair, justifying the use of the $\text{He}_2^{3+} \ ^2\Sigma_\text{g}^+$ Bethe logarithm values in the forthcoming computations.
 
\begin{figure}
  \centering
  \includegraphics[width=8.5cm]{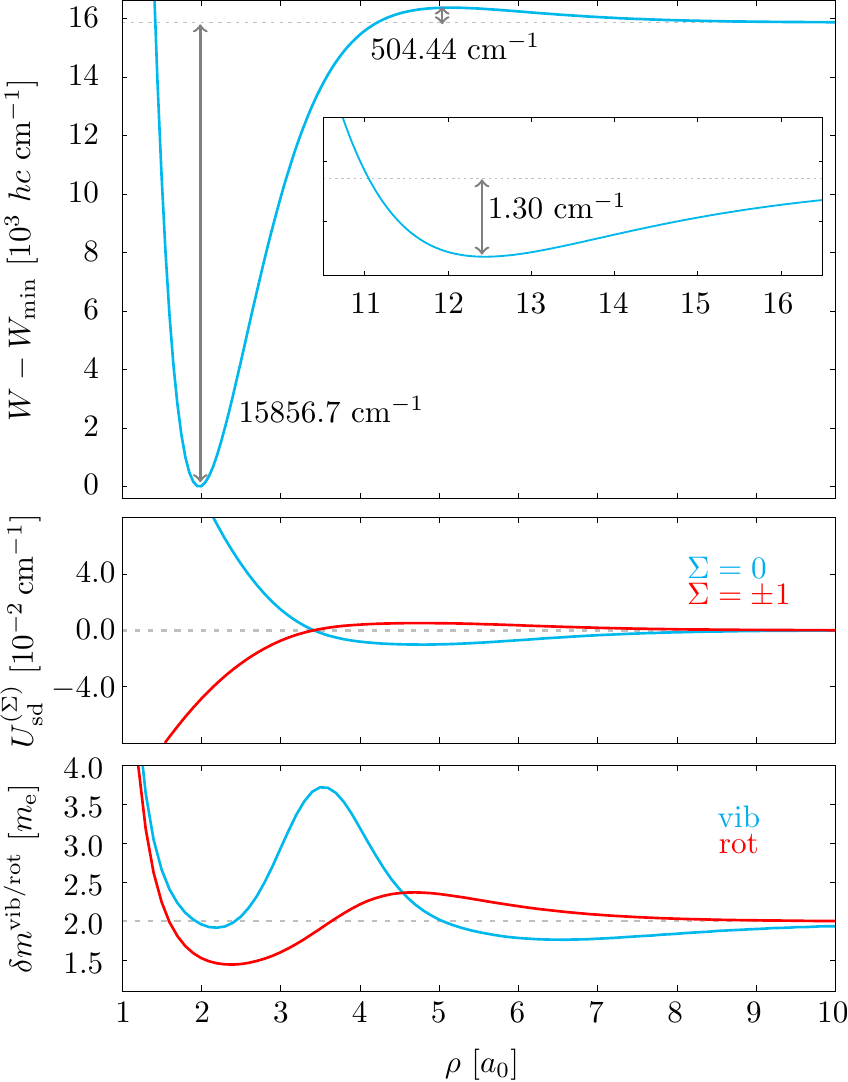}
  \caption{ %
     Top panel: corrected PEC of \atSup{} (with energy around the minimum 
     $W(\rho \approx 1.98 \, a_0)\approx-5.150\ 662\ 7 \, \Eh$  subtracted). 
     The dashed line represents the dissociation limit %
     $W(\rho\rightarrow \infty)= W[1 \, ^1S] + W[2 \, ^3S]\approx -5.078\ 414\ 1\ \Eh$. 
     Middle panel: spin-dependent correction to the PEC. 
     Bottom panel: non-adiabatic vibrational and rotational mass corrections. 
     All data used to prepare these figures (up to 100~$a_0$) are provided as \SM{}.
     \label{fig:acentroid}%
  }
\end{figure}

Using these expectation values, all corrections to the PEC can be calculated. The top panel of Fig.~\ref{fig:acentroid} shows the full, corrected potential energy curve $W$ summarizing all spin-independent and adiabatic effects; the spin-dependent contributions, $U^{(\Sigma)}_\text{sd}$, are shown in the middle panel (computed with the implementation according to Ref. \cite{paper3}).
The minimum of the PEC is $W(1.98 \, a_0)=-5.150 \, 662 \, 7
 \, \Eh$;
the potential energy barrier, $W(5.14 \, a_0)=-5.076 \, 115 \, 8
 \, \Eh$, and the second minimum, $W(12.45 \, a_0)= -5.078 \, 420 \, 1
 \, \Eh$, can be also seen in the figure (and the inset).
Curves of the computed vibrational and rotational mass corrections are displayed in the bottom panel; some details are given in the \SM{}.

\section{Solving the rovibronic problem and comparison with experimental data \label{sec:rovib}}
 
The following electron-spin-rotation-vibration Ansatz is used for the nuclear Schrödinger equation \cite{YuLoTeSt16,LBFi04book,BrCa03}
\begin{align}
  \Psi_{\text{a};J,M_J}(\br,\rho,\theta,\phi,\chi) = 
  \sum_{\Sigma=-S}^{+S}
  \sum_{k}c_{\Sigma,k} \, 
    \vphi_\text{a}^{(\Sigma)} \,  D^J_{M_J\Sigma} \, \frac{1}{\rho}g_k  \; ,
\end{align}
where $\vphi_\text{a}^{(\Sigma)}(\br;\brho)$ is the eigenfunction of $\hat{H}_\text{el}(\brho)$, the (normalized) $D^J_{M_J \Sigma}(\theta,\phi,\chi)$ Wigner $D$ matrix is the basis function for the rotation of the (diatomic) nuclear skeleton with the electron spin, and $\{g_k(\rho)\}$ are vibrational basis functions. 
The quantum numbers of the total rotational plus electron spin angular momenta, $J=0,1,2,\ldots$ and the corresponding laboratory-fixed $Z$ projection, $M_J=-J,\ldots,+J$, are exact quantum numbers of the theory; if no external fields are present, the rovibronic states are degenerate in $M_J$.

The quantum dynamics of the nuclei is described by a second-order effective rovibrational Hamiltonian~\cite{PaSpTe07,Ma18nonad,Ma18he2p,FeMa19HH,MaTe19}. Relativistic, QED and finite size effects are added to this Hamiltonian in the form of corrections to the adiabatic electronic energy and couplings between electronic states.
Without going into further details (described in Ref.~\cite{paper2}), the solution of the rovibrational problem for a well-separated $\Sigma_\text{g/u}$ electronic state of a homonuclear diatomic molecule boils down to solving the following radial equation for $g_k(\rho)$, 
\begin{align}
&E
  \sum_{k} c^{(\Sigma)}_{k} g_k= \nonumber \\
&\sum_{k}\Bigg\{
  c^{(\Sigma)}_{k}\Bigg[
   -\frac{\text{d}}{\text{d}\rho} \frac{1}{2\mu^\text{v}}\frac{\text{d}}{\text{d} \rho}
   +\frac{J(J+1)-\Sigma^2}{2\mu^\text{r} \rho^2} \nonumber \\
  &+\frac{S(S+1)-\Sigma^2}{2\mu^{\rm r}\rho^2}
      +W +U^{(\Sigma)}_{\text{rel+QED,sd}}\Bigg] \nonumber \\
&-\frac{c^{(\Sigma+1)}_{k}C_{J,\Sigma+1}^{-}C_{S,\Sigma+1}^{-}+c^{(\Sigma-1)}_{k}C_{J,\Sigma-1}^{+}C_{S,\Sigma-1}^{+}}{2\mu^{\rm r}\rho^2} 
  \Bigg\}g_k
\ ,
  \label{rovibeq}
\end{align}
where $C_{AB}^{\pm} = \sqrt{A(A+1)-B(B\pm 1)}$, and the corrected PEC $W$, its spin-dependent part $U^{(\Sigma)}_{\text{sd}}$ and the non-adiabatic mass corrections were all discussed previously.
Terms in the last line of Eq. (\ref{rovibeq}) couple states with different $\Sigma$ values.

The corrected PEC and the mass corrections together provide the input for solving the rovibrational problem. The radial equation, 
Eq.~(\ref{rovibeq}), was solved in a Discrete Variable Representation (DVR) \cite{LiCa00} using associated Laguerre polynomials of second order, over the range $\rho/a_0\in[1,10]$, on a grid of typical size $N_\text{grid}=800$ (the results reported in this paper were converged to $10^{-5}\ \cm$ with this grid). At the end, the Pauli-allowed rovibronic states are selected that are symmetric to the exchange of the two nuclei (the $^4$He nuclei being bosons); this leaves only odd values of the rotational quantum number. We found a total of 198 bound states with vibrational quantum numbers up to $v=12$. Below, we compare our computational results with the available experimental data; the computed energy level list of all 198 bound states found is deposited in the \SM{}.

\subsection{Ionization energy}
The ionization energy is computed as the difference of the $(v=0,N=1)$ rovibrational ground state energies of \XdSup\ He$_2^+$ and \atSup\ He$_2$:
\begin{align}
 E_{\text{X$^+$},01}&=-4.990\ 232\ 697(10) \, \Eh \ , \\
E_{\text{a},01}&=
-5.146\ 519\ 84 (100)\, \Eh \ ,
\end{align}
improved results for the former are taken from Ref.~\cite{paper-he2p}.  
The uncertainty of both values is dominated by the convergence error of the BO energy, which is a rigorous upper bound to the exact BO energy. We also note that the $E_{\text{a},01}$ value already includes the correction from the single-point further optimization of the BO energy (Table~\ref{tab:IP} \& \SM{}).
\begin{table}[]
\caption{%
    Ionization energy of \atSup\ He$_2$: 
    comparison of the computed and experimental values, 
    $\Delta\tilde\nu^{\text{X}^+\text{a}}_{01}=\tilde\nu^{\text{X}^+}_{01}-\tilde\nu^{\text{a}}_{01}$,
    corresponding to the \atSup(0,1)~$\rightarrow$~\XdSup(0,1) interval. 
    All values in $\text{cm}^{-1}$. The theoretical contributions and the deviation from experiment are displayed as $\Delta\tilde\nu_\text{part}$ and
    $\delta\tilde\nu
    =
    \Delta\tilde\nu^{\text{X}^+\text{a}}_{01,\text{Expt.}}
    -
    \Delta\tilde\nu^{\text{X}^+\text{a}}_{01,\text{Comp.}}$, respectively. 
    }
    \label{tab:IP}
    \centering
    \begin{tabular}{@{}l@{}d{6.6}d{5.5}d{6.5}@{}}
    \hline\hline\\[-0.35cm] &
    \multicolumn{1}{c}{$\Delta\tilde\nu_\text{part}$}
     
     & \multicolumn{1}{c}{$\Delta\tilde\nu^{\text{X}^+}_{01}$} & 
     \multicolumn{1}{c}{$\delta\tilde\nu$}
     \\[0.05cm]
     \hline\\[-0.35cm]     %
BO only      &34303.8923&	34\ 303.89(70)&		-2.685 \\
+ DBOC &	-4.2135&34\ 299.68(70)&		1.528 \\
+ nonad &	0.0531&34\ 299.73(70)&		1.475 \\
+ rel &	0.9175&34\ 300.65(70)&		0.558 \\
+ lQED &	-0.0925&34\ 300.56(70)&		0.650 \\
+ hQED + nuc &	-0.0018&34\ 300.56(70)&		0.652 \\ \hline\\[-0.35cm]
+ BO corr.$^\text{a}$        &0.5170&	34\ 301.07(25) &		0.139 \\ 
     \hline\\[-0.35cm]
       Expt.~\cite{SeJaCaMeScMe20}&  & 34\ 301.207\ 00(4)  & \\
     \hline\hline         
    \end{tabular}
    \begin{flushleft}
      $^\text{a}$  Rovibrational ground state energy of $\text{a} \, ^3\Sigma_\text{u}^+$ corrected with the single-point electronic energy difference corresponding to the $N_\text{b}=2500$ and 1500(PEC) basis sets (Table S2 of \SM).
    \end{flushleft}
\end{table}

By considering all corrections (Table~\ref{tab:IP}), the ionization energy is $34301.07(25) \ \text{cm}^{-1}$, to be compared with the experimental value, $\Delta\tilde\nu^{\text{X}^+\text{a}}_{01,\text{Expt.}}=34301.20700(4) \ \text{cm}^{-1}$~\cite{SeJaCaMeScMe20}. 
The deviation of the computed data from the experimental value is still dominated by the BO energy convergence error. 
Further computations can better converge the BO PEC, as shown for the $\rho=2~a_0$ geometry (single point) in the \SM{}. 
Since the PEC was generated in a smooth, consecutive fECG optimization sequence, the relative error of the PEC is much smaller and can be used for computing precise rovibrational intervals.

\subsection{Vibrational intervals}
Energy differences associated with the (Pauli-forbidden, hence fictitious) $(0,0)\rightarrow(v,0)$ vibrational excitations are displayed in Table~\ref{tab:vibspacing}. 
\begin{table}[]
  \caption{%
    Vibrational intervals of \atSup\ He$_2$, 
    $\Delta\tilde{\nu}_{v}=\tilde\nu_{v0}-\tilde\nu_{00}$, 
    and their deviation, 
    $\delta\tilde\nu=\Delta\tilde\nu_{v}^\text{Expt.}-\Delta\tilde\nu_{v}^\text{Comp.}$ 
    from the empirical values 
    compiled from experimental data \cite{FoBeCo98,BrGi71}.
     All quantities in $\cm$.
    The overall estimated uncertainty for the computed values listed in this table is 0.05~$\cm$.
    }
  \label{tab:vibspacing}
  \centering
   \begin{tabular}{@{}r @{\ \ \ }d{5.3}d{5.9}cc c@{}}
    \hline\hline\\[-0.4cm]
    &   
    \multicolumn{3}{c}{$\tilde\nu_{v0}-\tilde\nu_{00}$} & 
    \raisebox{-0.25cm}{$\delta\tilde\nu$} \\[-0.10cm]
    \cline{2-3}\\[-0.35cm]
    $v$  &  \multicolumn{1}{c}{Comp.}  & \multicolumn{2}{c}{Expt.}   &     \\
    \hline\\[-0.35cm]
   1 & 1732.19 & 1732.1615(23)^\text{a} & \cite{FoBeCo98} & $-$0.03 \\
   2 & 3386.54 & 3386.5023(60)^\text{a} & \cite{FoBeCo98} & $-$0.04 \\
   \hline \\[-0.35cm]
   3 & 4961.25 & 4961.12(3)^\text{b,c}   & \cite{BrGi71} & $-$0.1 \\
   4 & 6454.06 & 6453.88(6)^\text{b}   & \cite{BrGi71} & $-$0.2 \\
   5 & 7862.23 & 7861.98(16)^\text{b}  & \cite{BrGi71} & $-$0.2 \\
    \hline\hline\\[-0.6cm]     
  \end{tabular}
  \begin{flushleft}
  {%
    $^\text{a}$ Ref.~\cite{FoBeCo98},
    $^\text{b}$ Ref.~\cite{BrGi71}; in both cases, the `empirical' values were taken from the effective Hamiltonian fitted to the experimentally measured rovibrational transitions. \\
    $^\text{c}$ New experimental data and analysis~\cite{HoWiShBeMe25} indicate that the experiment-theory agreement may be better, within 0.03--0.07~$\cm$ (please see also text).
  }
  \end{flushleft}
\end{table}
The `experimental' values are obtained from the molecular constants of effective Hamiltonians fitted to 
rovibronic transitions involving \atSup\ states.

The agreement is very good between the computed and experimental values for the first few vibrational levels, while apparently less satisfactory for higher levels. This might be either due the non-uniformity of the BO PEC error propagating to the higher vibrational levels or due to the experimental values of Ref. \cite{BrGi71} not being accurate enough. New experimental data and analysis including the a$vN$ rovibrational levels with $v=3$ and $N=5,7,9,11$ suggest (through comparison of our work with data in the Supporting Information of Ref.~\cite{HoWiShBeMe25}) that our $v=3$~--~$v=0$ vibrational interval is accurate to 0.03--0.07~$\cm$. 

\begin{table}[]
  \caption{%
    Rotational intervals, $\Delta\tilde\nu_{vN}=\tilde\nu_{vN}-\tilde\nu_{v1}$, in $\cm$,
    within the $v=0,1$ vibrational bands of He$_2$~\atSup. 
    Deviation from experiment, 
    $\delta\nu=\Delta\tilde\nu_{vN}^{\text{Expt.}}-\Delta\tilde\nu_{vN}^\text{Comp.}$ 
    in $\cm$, is also shown. 
The overall estimated uncertainty for the computed values listed in this table is 0.001~$\cm$.
    } 
  \label{tab:rotinterval}
  \centering
  \begin{tabular}{@{}r@{\ \ } r@{\ \ } d{5.10}@{\ }l@{\ \ }d{2.6}@{}}
  \hline\hline\\[-0.35cm]
   & 
   \multicolumn{3}{c}{$\tilde\nu_{0N}-\tilde\nu_{01}$} & 
   \multicolumn{1}{c}{\raisebox{-0.25cm}{$\delta\tilde\nu$}}  \\[-0.10cm]
   \cline{2-4} \\[-0.35cm]
   $N$  &  
   \multicolumn{1}{c}{Comp.} & 
   \multicolumn{2}{c}{Expt.}  & \\
   \hline \\[-0.35cm]
   \multirow{2}{*}{3} &   \multirow{2}{*}{75.812\ 95}   &    75.812\ 994(11)  & $^\text{a}$ &  0.000\ 04 \\
    &     &    75.813\ 6(8)  & $^\text{b}$ &  0.000\ 6 \\
    \multirow{2}{*}{5} & \multirow{2}{*}{211.993\ 90}   &   211.993\ 994(13)  & $^\text{a}$ &  0.000\ 09 \\
     &     &   211.995\ 0(8)  & $^\text{b}$ &  0.001\ 1 \\
     \multirow{2}{*}{7} &  \multirow{2}{*}{408.060\ 83}   &   408.060\ 984(16)  & $^\text{a}$ &  0.000\ 15 \\
      &     &   408.061\ 4(8)  & $^\text{b}$ &  0.000\ 6 \\
      \multirow{2}{*}{9} &  \multirow{2}{*}{663.321\ 82}   &   663.322\ 009(19)  & $^\text{a}$ &  0.000\ 19 \\
       &    &   663.323\ 1(8)  & $^\text{b}$ &  0.001\ 3 \\
       11 &  976.879\ 69   &   976.880\ 9(8)  & $^\text{b}$ &  0.001\ 2  \\
13 & 1347.637\ 87   &  1347.639\ 6(8)  & $^\text{b}$ &  0.001\ 7  \\
15 & 1774.307\ 18   &  1774.307\ 2(8)  & $^\text{b}$ &  0.000\ 0  \\
17 & 2255.413\ 62   &  2255.413\ 3(8)  & $^\text{b}$ & -0.000\ 3 \\
19 & 2789.306\ 65   &  2789.305\ 6(8)  & $^\text{b}$ & -0.001\ 1
\\\hline\hline\\[-0.35cm]
   & 
   \multicolumn{3}{c}{$\tilde\nu_{1N}-\tilde\nu_{11}$} & 
   \multicolumn{1}{c}{\raisebox{-0.25cm}{$\delta\tilde\nu$}}  \\[-0.10cm]
   \cline{2-3} \\[-0.35cm]
   $N$  &  
   \multicolumn{1}{c}{Comp.} & 
   \multicolumn{2}{c}{Expt.} &  \\
   \hline\\[-0.35cm]
3   & 73.409\ 0    & 73.407\ 9(7)   & $^\text{c}$ & -0.001\ 1 \\
5   & 205.260\ 5   & 205.259\ 7(7)  & $^\text{c}$ & -0.000\ 8 \\
7   & 395.068\ 5   & 395.067\ 7(7)  & $^\text{c}$ & -0.000\ 8 \\
9   & 642.134\ 9   & 642.134\ 1(7)  & $^\text{c}$ & -0.000\ 8 \\
11  & 945.554\ 7   & 945.552\ 6(7)  & $^\text{c}$ & -0.002\ 1 \\
13  & 1304.220\ 7  & 1304.218\ 2(7) & $^\text{c}$ & -0.002\ 5 \\
\hline\hline
  \end{tabular}  
  \begin{flushleft}
    $^\text{a}$ Calculated as the centroid energy of the fine-structure components in Table~III of Ref.~\cite{WiHoMe25}. \\
    $^\text{b}$ Taken from Table~I of Ref.~\cite{SeJaMe16}. \\
    $^\text{c}$ Taken from Table~3.3 of Ref.~\cite{SemeriaPhD2020}, which reported data resulting from combination differences of rovibronic transitions from Refs.~\cite{RoBrBeBr88} and \cite{FoBeCo98}.
  \end{flushleft}  
\end{table}

\subsection{Rotational intervals}
Rotational intervals associated with $(0,1)\rightarrow(0,N)$ and $(1,1)\rightarrow(1,N)$ spacings are shown in Table~\ref{tab:rotinterval}. 
The agreement between theory and experiment is excellent for the 
$(0,1)\rightarrow(0,N)$ ($N=3,\ldots,9$) transitions where high-precision experimental data (of $1-2\cdot 10^{-5}$~$\cm$\ uncertainty) have recently become available \cite{WiHoMe25}.
Similarly, good agreement is observed for the rotational intervals 
$(0,1)\rightarrow(0,N)$ ($N=3,\ldots,19$) compared to Ref.~\cite{SeJaMe16} ($8\cdot 10^{-4}$~$\cm$\ experimental uncertainty),
as well as for the rotational intervals 
$(1,1)\rightarrow(1,N)$ ($N=3,\ldots,13$) in comparison with the data of Ref.~\cite{SemeriaPhD2020} (compiled from Refs.~\cite{RoBrBeBr88} and \cite{FoBeCo98}).
The computational uncertainty, a tentatively assigned overall value of $\sim10^{-3} \, \text{cm}^{-1}$, is dominated by the (much larger) PEC convergence error cancelling to a large extent between sublevels of the same vibrational band. For these intervals, we think that the most important missing physical effect in our theoretical model is the non-adiabatic-relativistic coupling.

\subsection{Fine-structure splittings}
Theoretical and experimental fine-structure splittings are compared in Table~\ref{tab:v0_QED_spinrot_off2}. The latter, `experimental' values are obtained from an effective Hamiltonian approach with parameters fitted to 
rovibronic excitation energies~\cite{BrCa03,FoBeCo98,SeJaClAgScMe18,WiHoMe25}
involving the (relevant) \atSup {} levels; see the Appendix and the `Expt.' columns of Table \ref{tab:spinspin_coeff} for the important formulae and the numerical values of the parameters, respectively. Although the $\{\gamma\}$ spin-rotation coupling constants are given in the Table for completeness, their values are set to zero when building and diagonalizing the effective Hamiltonian in order to be consistent with the theoretical computations, where this effect has not been not taken into account yet.
 
The agreement between computed and `experimental' splittings is excellent, the deviation being $250-500 \, \text{kHz}$ in almost all cases. This is a spectacular display of the importance of QED (in the form of anomalous magnetic moment corrections to $\hat{H}_{\text{SS,dp}}$, as shown in Eq.~(\ref{Hspindep})); omitting these corrections would introduce a mostly uniform error of $\sim3 \, \text{MHz}$. The remaining discrepancy is likely due to the neglected non-adiabatic-relativistic coupling corrections and higher relativistic corrections, whose effect can be roughly estimated by looking at the fine-structure splitting of hydrogen-like ions (\emph{e.g.} in Table~3 of Ref.~\cite{YeSh15}); these small corrections are to be investigated in future work.

Finally, for an overall, instead of a line-by-line, experiment-theory comparison, the rovibrational-fine-structure energy levels were used as input in the effective Hamiltonian to arrive at effective molecular parameters. Table~\ref{tab:spinspin_coeff} shows the  fitted values of the $B_i$ rotational and $\lambda_i$ ($i=0,1,2,3$) spin-spin coupling coefficients for the $v=0,1,2$ vibrational states and comparison with available experimental data. We used the same effective Hamiltonian expressions as Ref.~\cite{WiHoMe25}, and we see good agreement, especially for the spin-spin coupling parameters with the experimentally derived values. The computed higher rotational coefficients $B_2$ and $B_3$ are not reliable, their magnitude being well below the uncertainty of the computed rotational intervals. 
These coefficients should be recomputed in the future, once spin-rotation coupling is properly taken into account in the computation.

 \begin{table*}[]
    \caption{Fine-structure splitting of the rotational levels of He$_2$~\atSup\ in the $v=0,1,2$ vibrational bands, with QED corrections for the anomalous magnetic moment of the electron included. 
    The experimental (Expt.) results are constructed from the effective Hamiltonian treatment \cite{WiHoMe25} based on \cite{BrCoWaWa79,BrCa03} with rotational and spin-spin coupling parameters taken from the cited references, and the spin-rotational parameters set to zero.
    The theoretical uncertainty of the splittings is estimated to be $0.6\, \text{MHz}$ ($2\cdot10^{-5} \ \text{cm}^{-1}$).
 Deviation of experiment and computation, $\delta\nu=\nu^{\pm}_\text{Expt.}-\nu^{\pm}_\text{Comp.}$, is also shown.
    }
    \label{tab:v0_QED_spinrot_off2} 
    \centering
\scalebox{0.95}{%
    \begin{tabular}{@{}r@{\ \ \ \ }r@{\ \ \ \ }lcccclcccc@{}}
\hline\hline\\[-0.35cm]
 & &  
 \multicolumn{4}{c}{$J=N-1 \leftarrow J=N$} &&
 \multicolumn{4}{c}{$J=N+1 \leftarrow J=N$} \\
 \cline{3-6}\cline{8-11} \\[-0.35cm]
 $v$ & $N$ &  
 $\Delta\tilde\nu^{(-1)}_\text{Comp.}$ [cm$^{-1}$] & 
 $\Delta\tilde\nu^{(-1)}_\text{Expt.}$ [cm$^{-1}$] & &
 $\delta\nu^{(-1)}$ [MHz] &&
 $\Delta\tilde\nu^{(+1)}_\text{Comp.}$ [cm$^{-1}$] & 
 $\Delta\tilde\nu^{(+1)}_\text{Expt.}$ [cm$^{-1}$] & &
 $\delta\nu^{(+1)}$ [MHz] \\
 \hline\\[-0.35cm]
\multirow{11}{*}{0} %
&  1 &   0.073\ 284\ 8  &   0.073\ 302\ 40 & $^\text{a}$ &   0.53 &&  0.029\ 296\ 9  &   0.029\ 303\ 98  & $^\text{a}$  &   0.21 \\
&  3 &   0.043\ 908\ 8  &   0.043\ 919\ 37 & $^\text{a}$ &   0.32 &&  0.032\ 502\ 8  &   0.032\ 510\ 62  & $^\text{a}$  &   0.24 \\
&  5 &   0.040\ 518\ 6  &   0.040\ 528\ 31 & $^\text{a}$ &   0.29 &&  0.033\ 646\ 8  &   0.033\ 654\ 93  & $^\text{a}$  &   0.24 \\
&  7 &   0.039\ 085\ 0  &   0.039\ 094\ 21 & $^\text{a}$ &   0.28 &&  0.034\ 147\ 3  &   0.034\ 155\ 40  & $^\text{a}$  &   0.24 \\
&  9 &   0.038\ 190\ 2  &   0.038\ 198\ 60 & $^\text{a}$ &   0.25 &&  0.034\ 342\ 2  &   0.034\ 349\ 83  & $^\text{a}$  &   0.23 \\
& 11 &   0.037\ 496\ 6  &   0.037\ 506\ 22 & $^\text{b}$ &   0.29 &&  0.034\ 353\ 3  &   0.034\ 362\ 15  & $^\text{b}$  &   0.27 \\
& 13 &   0.036\ 882\ 9  &   0.036\ 892\ 63 & $^\text{b}$ &   0.29 &&  0.034\ 235\ 3  &   0.034\ 244\ 36  & $^\text{b}$  &   0.27 \\
& 15 &   0.036\ 294\ 9  &   0.036\ 304\ 78 & $^\text{b}$ &   0.30 &&  0.034\ 016\ 6  &   0.034\ 025\ 89  & $^\text{b}$  &   0.28 \\
& 17 &   0.035\ 704\ 8  &   0.035\ 714\ 90 & $^\text{b}$ &   0.30 &&  0.033\ 713\ 5  &   0.033\ 722\ 98  & $^\text{b}$  &   0.28 \\
& 19 &   0.035\ 097\ 1  &   0.035\ 107\ 33 & $^\text{b}$ &   0.31 &&  0.033\ 335\ 9  &   0.033\ 345\ 64  & $^\text{b}$  &   0.29 \\
& 21 &   0.034\ 462\ 2  &   0.034\ 472\ 65 & $^\text{b}$ &   0.31 &&  0.032\ 890\ 5  &   0.032\ 900\ 42  & $^\text{b}$  &   0.30 \\
\hline\\[-0.35cm]
\multirow{11}{*}{1} %
&  1 &   0.069\ 156\ 8  &   0.069\ 208\ 88 & $^\text{b}$ &   1.56 &&  0.027\ 647\ 1  &   0.027\ 667\ 38  & $^\text{b}$  &   0.61 \\
&  3 &   0.041\ 429\ 1  &   0.041\ 460\ 46 & $^\text{b}$ &   0.94 &&  0.030\ 667\ 8  &   0.030\ 690\ 26  & $^\text{b}$  &   0.67 \\
&  5 &   0.038\ 220\ 6  &   0.038\ 248\ 59 & $^\text{b}$ &   0.84 &&  0.031\ 738\ 9  &   0.031\ 761\ 71  & $^\text{b}$  &   0.68 \\
&  7 &   0.036\ 854\ 4  &   0.036\ 880\ 41 & $^\text{b}$ &   0.78 &&  0.032\ 198\ 8  &   0.032\ 221\ 17  & $^\text{b}$  &   0.67 \\
&  9 &   0.035\ 992\ 6  &   0.036\ 016\ 78 & $^\text{b}$ &   0.73 &&  0.032\ 366\ 3  &   0.032\ 387\ 77  & $^\text{b}$  &   0.64 \\
& 11 &   0.035\ 316\ 6  &   0.035\ 339\ 01 & $^\text{b}$ &   0.67 &&  0.032\ 356\ 2  &   0.032\ 376\ 55  & $^\text{b}$  &   0.61 \\
& 13 &   0.034\ 711\ 9  &   0.034\ 732\ 53 & $^\text{b}$ &   0.62 &&  0.032\ 220\ 4  &   0.032\ 239\ 27  & $^\text{b}$  &   0.57 \\
& 15 &   0.034\ 127\ 4  &   0.034\ 146\ 17 & $^\text{b}$ &   0.56 &&  0.031\ 985\ 3  &   0.032\ 002\ 74  & $^\text{b}$  &   0.52 \\
& 17 &   0.033\ 536\ 7  &   0.033\ 553\ 72 & $^\text{b}$ &   0.51 &&  0.031\ 666\ 4  &   0.031\ 682\ 30  & $^\text{b}$  &   0.48 \\
& 19 &   0.032\ 925\ 0  &   0.032\ 940\ 42 & $^\text{b}$ &   0.46 &&  0.031\ 273\ 0  &   0.031\ 287\ 44  & $^\text{b}$  &   0.43 \\
& 21 &   0.032\ 283\ 3  &   0.032\ 297\ 39 & $^\text{b}$ &   0.42 &&  0.030\ 811\ 1  &   0.030\ 824\ 35  & $^\text{b}$  &   0.40 \\
\hline\\[-0.35cm]
\multirow{11}{*}{2}
&  1 &   0.064\ 956\ 5  &   0.064\ 983\ 75 & $^\text{c}$ &   0.82 &&  0.025\ 968\ 4  &   0.025\ 979\ 24  & $^\text{c}$  &   0.33 \\
&  3 &   0.038\ 905\ 9  &   0.038\ 922\ 16 & $^\text{c}$ &   0.49 &&  0.028\ 800\ 5  &   0.028\ 812\ 55  & $^\text{c}$  &   0.36 \\
&  5 &   0.035\ 881\ 9  &   0.035\ 896\ 81 & $^\text{c}$ &   0.45 &&  0.029\ 797\ 2  &   0.029\ 809\ 60  & $^\text{c}$  &   0.37 \\
&  7 &   0.034\ 583\ 8  &   0.034\ 598\ 23 & $^\text{c}$ &   0.43 &&  0.030\ 215\ 3  &   0.030\ 227\ 92  & $^\text{c}$  &   0.38 \\
&  9 &   0.033\ 755\ 0  &   0.033\ 769\ 21 & $^\text{c}$ &   0.43 &&  0.030\ 354\ 3  &   0.030\ 367\ 14  & $^\text{c}$  &   0.38 \\
& 11 &   0.033\ 096\ 2  &   0.033\ 110\ 52 & $^\text{c}$ &   0.43 &&  0.030\ 322\ 2  &   0.030\ 335\ 29  & $^\text{c}$  &   0.39 \\
& 13 &   0.032\ 499\ 8  &   0.032\ 514\ 66 & $^\text{c}$ &   0.45 &&  0.030\ 167\ 2  &   0.030\ 180\ 96  & $^\text{c}$  &   0.41 \\
& 15 &   0.031\ 917\ 7  &   0.031\ 933\ 58 & $^\text{c}$ &   0.48 &&  0.029\ 914\ 4  &   0.029\ 929\ 34  & $^\text{c}$  &   0.45 \\
& 17 &   0.031\ 325\ 0  &   0.031\ 342\ 69 & $^\text{c}$ &   0.53 &&  0.029\ 578\ 1  &   0.029\ 594\ 86  & $^\text{c}$  &   0.50 \\
& 19 &   0.030\ 707\ 7  &   0.030\ 728\ 18 & $^\text{c}$ &   0.62 &&  0.029\ 167\ 0  &   0.029\ 186\ 45  & $^\text{c}$  &   0.58 \\
& 21 &   0.030\ 057\ 1  &   0.030\ 081\ 70 & $^\text{c}$ &   0.74 &&  0.028\ 686\ 4  &   0.028\ 709\ 93  & $^\text{c}$  &   0.70 \\
\hline\hline
    \end{tabular}
}
\begin{flushleft}
{\footnotesize%
  $^\text{a}$~Ref.~\cite{WiHoMe25}.  
}
\\
{\footnotesize%
  $^\text{b}$~Ref.~\cite{SeJaClAgScMe18}.  
}
\\
{\footnotesize%
  $^\text{c}$~Ref.~\cite{FoBeCo98}.
}
\end{flushleft}
\end{table*}

\begin{table*}[]
 \caption{%
 Condensed overview of the effective spin-spin couplings, $\lambda_0,\lambda_1,\lambda_2,\lambda_3$, modelling the fine-structure splittings (Table~\ref{tab:v0_QED_spinrot_off2}) through
 the effective Hamiltonian expressions (Appendix) fitted to the $N=1,...,21$ and $v=0,1,2$ energy levels; all values in $\text{cm}^{-1}$. 
 The same expressions were used for the computed (Comp.) data as for the most recent experimental (Expt.) values \cite{WiHoMe25}. 
 The electron-spin rotation coupling effect was not included in the computations, hence the $\{\gamma\}$ values are zero. The experimental $\{\gamma\}$ coefficients are shown only for the sake of completeness, but they were taken to be zero when calculating the 
 `Expt.' columns of Table~\ref{tab:v0_QED_spinrot_off2}.  
 }
    \label{tab:spinspin_coeff}
    \centering
\scalebox{0.95}{%
    \begin{tabular}{@{}l @{\ }c@{\ } d{4.8} d{2.12}  @{\ }c@{\ }  d{4.8} d{2.12}  @{\ }c@{\ }  d{4.8}d{2.12} @{}}
\hline\hline\\[-0.35cm]
 && 
 \multicolumn{2}{c}{$v=0$} &&
 \multicolumn{2}{c}{$v=1$} && 
 \multicolumn{2}{c}{$v=2$} \\
 \cline{3-4}
 \cline{6-7}
 \cline{9-10} \\[-0.35cm]
 &&
 \multicolumn{1}{c}{Comp.} & \multicolumn{1}{c}{Expt. \cite{WiHoMe25}} &&
 \multicolumn{1}{c}{Comp.} & \multicolumn{1}{c}{Expt. \cite{SeJaClAgScMe18}}  && 
 \multicolumn{1}{c}{Comp.} & \multicolumn{1}{c}{Expt. \cite{FoBeCo98}} \\
 \hline\\[-0.35cm]
 $B_0$                   &&   7.58851 &  7.5891610(10)  &&  7.34934 &  7.348742(32)  &&  7.10130 &  7.101747(120) \\
 $B_1\cdot10^{4}$        &&   5.451 &  5.61844(25)    &&  5.633 &  5.65381(178)  &&  5.554 &  5.7439(70) \\
 $B_2\cdot10^{8}$~$^\ast$        &&  [-4.93] &  3.395(17)      &&  [2.66] &  2.837(30)     && [-0.35] &  3.312(114) \\
 $B_3\cdot10^{12}$~$^\ast$       && [113.3] & -3.480          && [23.8] & \multicolumn{1}{c}{$-$} && [63.1] & \multicolumn{1}{c}{$-$} \\
 \hline\\[-0.35cm]
 $\lambda_0\cdot10^{2}$  &&  -3.66556 & -3.66643736(48) && -3.45918 & -3.4603756(17) && -3.24920 & -3.25056(86) \\
 $\lambda_1\cdot10^{6}$  &&   6.5904 &  6.58706(44)    &&  6.7148 &  6.7190(27)    &&  6.8566 &  6.864(22) \\
 $\lambda_2\cdot10^{10}$ &&  -1.560 & -1.558(18)      && -1.333 & -1.44(7)       && -1.022 & -1.595 \\
 $\lambda_3\cdot10^{14}$ &&   4.38 &  4.47(17)       &&  5.18 &  5.57(53)      &&  6.28 &  4.65\\
 \hline\\[-0.35cm]
 $\gamma_0\cdot10^{5}$   && 0          & -8.07931(39)    && 0         & -7.60693(57)   && 0         & -7.1466(67) \\
 $\gamma_1\cdot10^{8}$   && 0          &  2.2843(20)     && 0         &  2.171(2)      && 0         & 1.946(32) \\
 $\gamma_2\cdot10^{10}$  && 0          & -1.979(21)      && 0         & -1.78(2)       && 0         & \multicolumn{1}{c}{$-$} \\
%

\hline\hline
    \end{tabular}
}
\begin{flushleft}
  $^\ast$~$B_2$ and $B_3$ were sensitive to the fitting details of the computed data, and are not reliable; they were included only to be fully consistent with the procedure of the experimental work \cite{WiHoMe25}.
\end{flushleft}
\end{table*}

\begin{figure}
  \centering
  \includegraphics[width=8.5cm]{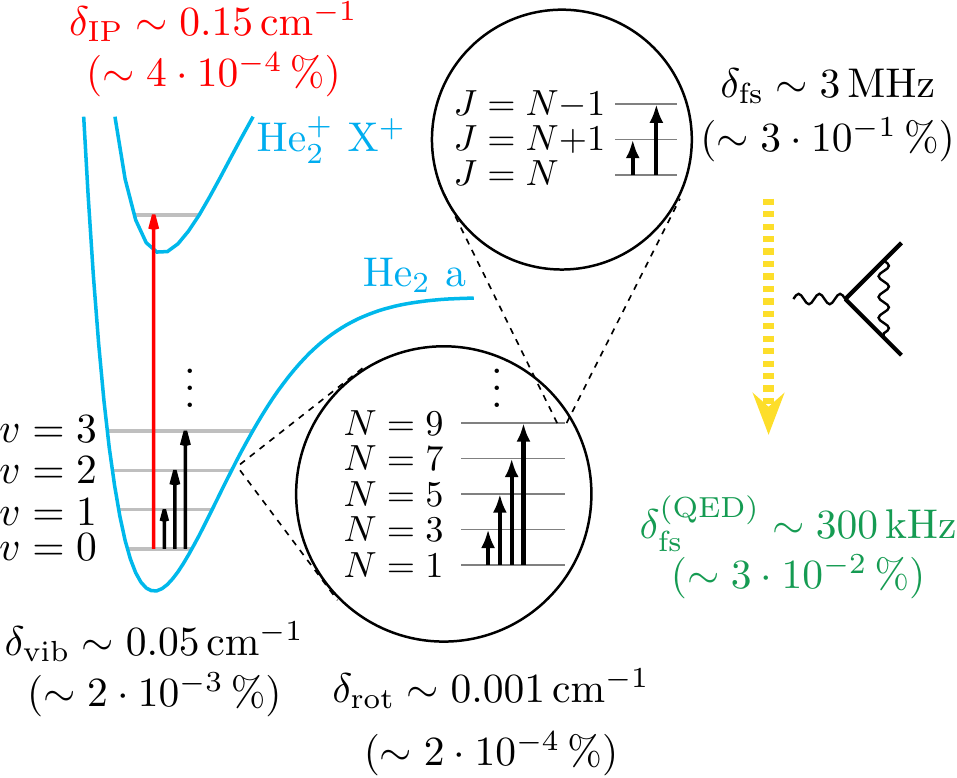}
  \caption{%
    A schematic depiction of the different energy scales accounted for in the \emph{ab initio} He$_2$ \atSup\ model developed in this work (not to scale). The typical absolute deviations from experimental values are shown as $\delta$ (the corresponding relative errors given in parentheses). \label{fig:multiscale}
  }
\end{figure}

\section{Summary, conclusion, outlook \label{sec:conclusion}}
\noindent%
This work reported the vibration-rotation-fine structure (Fig.~\ref{fig:multiscale}) of the lowest, \atSup{} electronic state of triplet He$_2$ with an unprecedented accuracy.
The computations were performed in the non-relativistic BO framework with relativistic, QED and non-adiabatic effects accounted for by perturbation theory.
The computation of the BO PEC in an explicitly correlated variational framework led to sub-ppm relative precision, which allowed us to study the effect of small corrections, \emph{e.g.,} dressed rotational masses (accounting for the effect of distant electronic states) and fine-structure splittings caused by magnetic spin-spin couplings, QED corrections of the latter being necessary to achieve good agreement with experiment. 

The various transitions in the molecule and the associated physical effects cover a range of 9 orders of magnitude in energy (Fig.~\ref{fig:multiscale}) from the $\sim10^4 \, \text{cm}^{-1}$ of ionization through $\sim10^3 \, \text{cm}^{-1}$ and $10^2 \, \text{cm}^{-1}$ of vibrational and rotational excitations to the $\sim10^{-2} \, \text{cm}^{-1}$ of the fine-structure splitting, the electron's anomalous magnetic moment correction to the latter appearing at the order of $10^{-5}$~$\cm$; all seen both in our computations and in the available experimental data. 
The scales of these transitions are set by the small perturbation parameters $(m_\text{el}/M_\text{nuc})^{1/2}$, $Z\alpha$, and $\alpha$ governing finite nuclear mass and relativistic \& QED effects, respectively. We relied on the separation of the different scales of the quantum mechanical molecular motions, which was essential for achieving quantitative agreement with experiments.

The detailed understanding of the rovibrational level structure of triplet He$_2$ may help in finding ionization pathways for more efficient generation of He$_2^+$, relevant for precision spectroscopy measurements. The excellent results achieved for the \atSup\ state open the route towards the higher-excited states of He$_2$ such as \btPg,  \ctSgp\ (and their singlet `counterparts' $\text{B} \, ^1\Pi_\text{g}$ and $\text{C} \, ^1\Sigma_\text{g}^+$), which introduce a new layer of complexity in the form of nonadiabatic and relativistic couplings among the electronic states, with interesting consequences for rovibrational quantum dynamics.

\vspace{0.25cm}
\section{Acknowledgement}
\noindent Financial support of the European Research Council through a Starting Grant (No.~851421) is gratefully acknowledged. We also thank the Momentum (Lendület) Programme of the Hungarian Academy of Sciences (LP2024-15/2024).
PJ acknowledges the János Bolyai Research Scholarship of the Hungarian Academy of Sciences (BO/285/22).
We thank DKF for awarding us access to the Komondor computing facility based in Hungary.
We thank Frédéric Merkt, Maximilian Holdener and Vincent Wirth for discussions regarding their recently published experimental results \cite{HoWiShBeMe25,WiHoMe25}.

\section*{Supporting Information Available}
\noindent %
The Supporting Information contains (a) further theoretical, computational details, and convergence tests; (b) the computed datasets and Wolfram Mathematica files used to compute the rovibrational-fine-structure levels.

%

\section*{Appendix: the effective Hamiltonian of fine-structure levels}
The rotational and fine-structure levels within a given vibrational band of an electronic state of $^3\Sigma$ symmetry are described by the effective Hamiltonian \cite{BrCa03}
 \begin{equation}
  \hat{H}_{\text{eff}}=\hat{B} \, \hat{\boldsymbol{N}}^2+\left[\hat{\lambda},\hat{S}_z^2-\frac{1}{3}\hat{\boldsymbol{S}}^2\right]_++\hat{\gamma} \, \hat{\boldsymbol{N}}\cdot\hat{\boldsymbol{S}} \ ,
 \end{equation}
 where 
 \begin{align}
  \hat{B}&=B_0- B_1 \, \hat{\boldsymbol{N}}^2+B_2(\hat{\boldsymbol{N}}^2)^2+B_3(\hat{\boldsymbol{N}}^2)^3+... \\
  \hat{\lambda}&=\lambda_0+\lambda_1 \, \hat{\boldsymbol{N}}^2+\lambda_2(\hat{\boldsymbol{N}}^2)^2+\lambda_3(\hat{\boldsymbol{N}}^2)^3+... \\
  \hat{\gamma}&=\gamma_0+\gamma_1 \, \hat{\boldsymbol{N}}^2+\gamma_2(\hat{\boldsymbol{N}}^2)^2+... \ .
 \end{align}
 The choice of the above effective Hamiltonian assumes a Hund $b$-type coupling between nuclear rotational, electronic orbital and electronic spin angular momenta ($\hat{\boldsymbol{R}}$, $\hat{\boldsymbol{L}}$ and $\hat{\boldsymbol{S}}$, respectively), meaning $\hat{\boldsymbol{J}}=(\hat{\boldsymbol{L}}+\hat{\boldsymbol{R}})+\hat{\boldsymbol{S}}=\hat{\boldsymbol{N}}+\hat{\boldsymbol{S}}$.
 The following energy levels are obtained from $\hat{H}_{\text{eff}}$:
 \begin{widetext}
 \begin{align}
  E_{N,J}=
  \left\{
                \begin{array}{ll}
                  \displaystyle\alpha_{N,J}\ \ \ \ \ \ \ \ \ \ \ \ \ \ \ \ \ \ \ \ \ \ \ \ \ \ \ \ \ \ \ \ \ \ \ \ \ \ \ \ \ \ \ \ \ \ \ \ \ \ \ \ \ \ \ \ \ \ \ \ \text{if $N=J$ or $J=0$,}\\
                  \displaystyle\frac{\alpha_{J+1,J}+\alpha_{J-1,J}\pm\sqrt{(\alpha_{J+1,J}-\alpha_{J-1,J})^2+4\beta_J^2}}{2} \ \ \ \ \text{if $N=J\pm1$ and $J\neq0$\ ,}
                \end{array}
              \right.
              \label{Heffeival}
 \end{align}
 where
 \begin{align}
 \alpha_{N,J}&=
 \left\{
                \begin{array}{ll}
                  \displaystyle B_N \, N(N+1)-\frac{2}{3}\lambda_N\frac{N}{2N+3}+\gamma_N \, N \ \ \ \  \ \ \ \ \ \ \ \text{if $J=N+1$}\\
                  \\
                 \displaystyle B_N \, N(N+1)+\frac{2}{3}\lambda_N-\gamma_N \ \ \ \ \ \ \ \ \ \ \ \ \ \ \ \ \ \ \ \ \ \ \ \ \  \text{if $J=N$} \\
                 \\
                 \displaystyle B_N \, N(N+1)-\frac{2}{3}\lambda_N\frac{N+1}{2N-1}-\gamma_N \, (N+1) \ \ \ \ \text{if $J=N-1$} \\
                \end{array}
              \right.
 \\
 \beta_{J}&=\frac{\sqrt{J(J+1)}}{2J+1}\left(\lambda_{J+1}+\lambda_{J-1}\right) \ ,
 \end{align}
 and
 \begin{align}
  B_N&=B_0- B_1 \, N(N+1)+B_2 \, [N(N+1)]^2+B_3 \, [N(N+1)]^3+... \\
  \lambda_N&=\lambda_0+\lambda_1 \, N(N+1)+\lambda_2 \, [N(N+1)]^2+\lambda_3 \, [N(N+1)]^3+... \\
  \gamma_N&=\gamma_0+\gamma_1 \, N(N+1)+\gamma_2 \, [N(N+1)]^2+... \ .
 \end{align}
 \end{widetext}
 Secs. 8.3.2, 9.6.1 and Appendix 8.3 of Ref. \cite{BrCa03} provide a detailed discussion on the effective Hamiltonian and the computation of matrix elements; we also cite Ref.~\cite{WiHoMe25} for the final expressions of the energy levels.
The actual expressions are reiterated here for completeness.

\clearpage
\onecolumngrid

\clearpage
\begin{center}
{\large
\textbf{Supporting Information}
}\\[0.25cm]
{\large
\textbf{Rovibrational computations for the He$_2$ \atSup\ state including non-adiabatic, relativistic, and QED corrections}
} \\[0.5cm]

Ádám Margócsy,$^1$ Balázs Rácsai,$^1$ Péter Jeszenszki,$^1$ Edit Mátyus$^{1,\ast}$ \\
\emph{$^1$~MTA–ELTE Lendület `Momentum' Molecular Quantum electro-Dynamics Research Group,
Institute of Chemistry, Eötvös Loránd University, Pázmány Péter sétány 1/A, Budapest, H-1117, Hungary} \\
$^\ast$ edit.matyus@ttk.elte.hu
~\\[0.15cm]
(Dated: 24 February 2026)
\end{center}

\renewcommand{\thepage}{S\arabic{page}}
\setcounter{page}{1}

\setcounter{section}{0}
\renewcommand{\thesection}{S\arabic{section}}
\setcounter{subsection}{0}
\renewcommand{\thesubsection}{S\arabic{section}.\arabic{subsection}}

\setcounter{equation}{0}
\renewcommand{\theequation}{S\arabic{equation}}

\setcounter{table}{0}
\renewcommand{\thetable}{S\arabic{table}}

\setcounter{figure}{0}
\renewcommand{\thefigure}{S\arabic{figure}}

\vspace{0.5cm}

\begin{table}[!h]
  \caption{Numerical values of the physical constants and conversion factors used in the present computations (taken from CODATA~2022~\cite{codata22}). 
  \label{tab:constants}   
  } 
  \centering  
  \begin{tabular}{@{}lr@{}}
  $\alpha$ & $7.2973525643(11)\cdot10^{-3}$ \\
     $M(^4\text{He}) \, / \, m_\text{e}$  & $7.29429954171(17)\cdot10^{+3}$  \\
     $\Eh \, / \, (hc \, \text{cm}^{-1})$ & $2.1947463136314(24)\cdot10^{+5}$  \\
  $\Eh \, / \, (h \, \text{MHz})$ & $6.579683920 4999(72)\cdot10^{+9}$  \\
  ${\cal{R}}(^{4}\text{He}^{2+}) / \lambdabar_\text{C}$ & $4.3467(54)\cdot10^{-3}$
  \end{tabular}
\end{table}

\section{Spin eigenfunctions}\label{sec:SpinFunctions}
\noindent
When acting on some state with well-defined $N_{\alpha}$ and $N_{\beta}$ (that is, well-defined $\Sigma=(N_\alpha-N_\beta)/2$), the action of $\hat{S}^2$ takes a simple form,
\begin{align}
  \hat{S}^2|N_\alpha,N_\beta\rangle
  =&
  \left[\left(\frac{N_\alpha-N_\beta}{2}\right)^2+\frac{N_\alpha+N_\beta}{2}\right]|N_\alpha,N_\beta\rangle +\hat{P}_{\alpha\beta}|N_\alpha,N_\beta\rangle \ ,
  \label{S2simple}
\end{align}
where $\hat{P}_{\alpha\beta}$ interchanges \emph{single} $\alpha$-$\beta$ labels in every possible way. Using this relation, we constructed the matrix of $\hat{S}^2$ in a block-diagonal form for each $\Sigma$ block. The orthonormal spin eigenstates have the form,
\begin{equation}
  |{\Theta_{S,\Sigma}}\rangle
  =
  \sum_{\substack{m_{s_1},...,m_{s_{\nel}} \\ 
  \sum_k m_{s_{k}}=\Sigma}}Q^{S,\Sigma}_{m_{s_1}...m_{s_{\nel}}}|\eta_{{s_1}}\eta_{m_{s_2}}...\eta_{m_{s_{\nel}}}\rangle \ ,
\end{equation}
so that $\hat{S}_z|\Theta_{{S,\Sigma}}\rangle=\Sigma|\Theta_{S,\Sigma}\rangle$ and $\hat{S}^2|\Theta_{S,\Sigma}\rangle=S(S+1)|\Theta_{S,\Sigma}\rangle$;  $m_{s_k}=\pm1/2$ is the spin projection value of a single-electron spin function represented by $\eta_{m_{s_k}}$. 
Note that $-$ due to degeneracy $-$, multiple equivalent forms can be found for the eigenvectors with the same $S$ and $\Sigma$;
these can always be converted into each other by suitable unitary linear combinations.
  
One can form $2^4=16$ different products of spin functions, from which, using Eq. (\ref{S2simple}), the following three sets of triplet spin eigenstates are found:
\begin{align}
   |\Theta_{1,-1}\rangle&=\frac{1}{\sqrt{12}}\big( |{\beta\beta\beta\alpha}\rangle + |{\beta\beta\alpha\beta}\rangle -
                                          3 |{\beta\alpha\beta\beta}\rangle + |{\alpha\beta\beta\beta}\rangle  \big) \\
   |\Theta_{1,0}\rangle&=\frac{1}{\sqrt{2}}\big( |{\alpha\alpha\beta\beta}\rangle - |{\beta\beta\alpha\alpha}\rangle  \big) \\
   |\Theta_{1,+1}\rangle&=\frac{1}{\sqrt{12}}\big( |{\alpha\alpha\alpha\beta}\rangle + |{\alpha\alpha\beta\alpha}\rangle -
                                           3|{\alpha\beta\alpha\alpha}\rangle + |{\beta\alpha\alpha\alpha}\rangle  \big) \\ 
   |\Theta'_{1,-1}\rangle&=\frac{1}{\sqrt{6}}\big( |{\beta\beta\beta\alpha}\rangle -2 |{\beta\beta\alpha\beta}\rangle + 
                                            |{\alpha\beta\beta\beta}\rangle  \big) \\ 
   |\Theta'_{1,0}\rangle&=\frac{1}{\sqrt{2}}\big( |{\alpha\beta\alpha\beta}\rangle - |{\beta\alpha\beta\alpha}\rangle  \big) \\
   |\Theta'_{1,+1}\rangle&=\frac{1}{\sqrt{6}}\big( |{\alpha\alpha\alpha\beta}\rangle -2 |{\alpha\alpha\beta\alpha}\rangle + {\beta\alpha\alpha\alpha}\rangle  \big) \\ 
   |\Theta''_{1,-1}\rangle&=\frac{1}{\sqrt{2}}\big( |{\beta\beta\beta\alpha}\rangle - |{\alpha\beta\beta\beta}\rangle  \big) \\
   |\Theta''_{1,0}\rangle&=\frac{1}{\sqrt{2}}\big( |{\alpha\beta\beta\alpha}\rangle - |{\beta\alpha\alpha\beta}\rangle  \big) \\
   |\Theta''_{1,+1}\rangle&=\frac{1}{\sqrt{2}}\big( |{\alpha\alpha\alpha\beta}\rangle - |{\beta\alpha\alpha\alpha}\rangle  \big) 
\end{align}
In the computational practice, the linear combination of the three triplet spin functions is used for a given $\Sigma$. Restricting ourselves to real-valued parameters, the expansion is determined by $2$ `angle parameters':
\begin{equation}
 C(\boldsymbol{\theta})=\sin(\theta)\cos(\phi) \ \ , \ \ C'(\boldsymbol{\theta})=\sin(\theta)\sin(\phi) \ \ , \ \ C''(\boldsymbol{\theta})=\cos(\theta) \ .
\end{equation}
This linear combination is reexpressed in terms of spin function products \cite{SuVaBook98}
\begin{align}
 \chi_{S,\Sigma}(\boldsymbol{\theta})
 &=
 C(\boldsymbol{\theta}) \, \Theta_{S,\Sigma}+C'(\boldsymbol{\theta}) \, \Theta'_{S,\Sigma}+C''(\boldsymbol{\theta}) \, \Theta''_{S,\Sigma} \nonumber \\
 &=
 \sum_{\substack{m_{s_1},...,m_{s_4} \\
   \Sigma m_{s_k}=\Sigma}}K^{S,\Sigma}_{m_{s_1},m_{s_2},m_{s_3},m_{s_4}}(\boldsymbol{\theta}) \, [\eta_{m_{s_1}}\otimes \eta_{m_{s_2}}\otimes\eta_{m_{s_3}}\otimes \eta_{m_{s_4}}] \ .
\end{align}
It was shown in Ref.~\citenum{CeRy93} that a single spin function is sufficient from the relevant subspace; nevertheless, we retain all possible functions which (as additional free parameters) may help improve the BO energy convergence.

\section{On the numerical solution of the electronic equation: convergence of the electronic energy and its corrections}
\subsection{Born-Oppenheimer energy curve}
\noindent
The electronic Schrödinger equation
\begin{equation}
 \hat{H}_{\text{el}}(\brho)\vphi_\text{a}^{(\Sigma)}(\br;\brho)=U_\text{a}(\rho)\vphi_\text{a}^{(\Sigma)}(\br;\brho)
 \label{Sch_electronic}
\end{equation}
is solved by parametrizing $\vphi_\text{a}^{(\Sigma)}(\br;\brho)$ as a truncated expansion of $N_\text{b}$ symmetry-adapted floating explicitly correlated Gaussians~(fECG-s)~\cite{SuVaBook98,MaRe12,MiBuHoSuAdCeSzKoBlVa13}:
\begin{equation}
 \vphi(\br)
 =
 \sum_{\nu=1}^{N_\text{b}}c_{\nu} \, 
   {\cal{P}}_{G} \, {\cal{A}}_S
   \left\lbrace
   \exp\left[%
    -(\br-\bs_\nu)^\tT \underline{\bA}_\nu (\br-\bs_\nu)
  \right]\chi_{S,\Sigma}(\boldsymbol{\theta}_\nu) 
  \right\rbrace
  \ ,
 \label{ECGexpansion}
\end{equation}
where $\underline{\bA}_\nu=\bA_\nu\otimes\boldsymbol{I}_3$ with $\bA_\nu\in\mathbb{R}^{\nel\times \nel}$ is a symmetric positive-definite matrix, $\boldsymbol{I}_3\in\mathbb{R}^{3\times3}$ is the three-dimensional unit matrix, and $\bs_\nu\in\mathbb{R}^{3\nel}$ is the center of the fECG in the configuration space.
The $\chi_{S,\Sigma}$ spin part is a linear combination of all $N_\text{s}(\nel,S)$ spin eigenfunctions with given $S,\Sigma$, rewritten as an expansion over product spin functions (\emph{vide supra}). The ${\cal{A}}_S$ and ${\cal{P}}_{G}$ projection operators implement the Pauli principle and the point-group symmetry (in our case, $\Sigma_{\text{u}}^+$ of $D_{\infty \text{h}}$). 

Substituting Eq.~(\ref{ECGexpansion}) into Eq.~(\ref{Sch_electronic}) yields a generalized matrix eigenvalue equation $\boldsymbol{H}\boldsymbol{c}=E\boldsymbol{S}\boldsymbol{c}$ 
with $H_{\nu\tau}=\langle\phi_\nu|\hat{H}_\text{el}|\phi_\tau\rangle$ and $S_{\nu\tau}=\langle\phi_\nu|\phi_\tau\rangle$ (both calculable analytically). This is a linear variational problem for $c_\nu$ (found via diagonalization), and a nonlinear one for $\bA_\nu,\bs_\nu$, $\bos{\theta}_\nu$; the latter are generated according to the stochastic variational approach \cite{SuVaBook98} and optimized with the Powell method \cite{Po04}. The electronic energy of few-particle systems can be converged to high precision with this technique~\cite{MiBuHoSuAdCeSzKoBlVa13}.

The fECG optimization has been performed with our in-house developed computer program, QUANTEN. The $U_\text{a}(\rho)$ potential energy curve (PEC) was computed at several points over the range $\rho/a_0\in[1.00,100.00]$ starting from a thoroughly optimized point at $2 \, a_0$, and always using the (rescaled) wave function parameters of the previous point as an initial guess for the next one \cite{CeRy95,FeMa19HH,FeKoMa20,FeMa22h3}. The curve was computed with step size $0.05 \, a_0$ for $\rho/a_0\in[1.0,10.00]$, $0.10 \, a_0$ for $\rho/a_0\in[10.00,16.50]$ and $1.00 \, a_0$ for $\rho/a_0\in[17.0,100.0]$. 

The optimization of the $N_\text{b}=1500$ basis for the $2 \ a_0$ point took roughly 150 000 CPU hours. Developing the BO PEC starting from this point on intervals $[1 \ a_0,17 \ a_0]$ and $[17 \ a_0,100 \ a_0]$ took approximately 150 000 CPU hours and 50 000 CPU hours, respectively (fewer points were computed in the long-range interval, and they required fewer optimization cycles).

During (and after) the PEC generation, we continued the (single-point, $2\ a_0$) fECG optimization for $N_\text{b}=1500$, and further increasing the basis size to $N_\text{b}=2500$ revealed an improvement of roughly $2 \, \mu\Eh$ at $\rho=2 \, a_0$ compared to the energy that was used to develop the PEC (see the last rows of the fECG panel of Table \ref{tab:BOconv}). 
%

The $N_\text{b}=1000,1500,2000,2500$ energies were fitted to the model 
$E=E_\infty+a \, N_\text{b}^{-k}$, from which the energy was extrapolated to the complete basis set limit, $E_\infty\approx-5.151\ 125\ 050\ \Eh$. 
Based on this, an error of roughly $\sim3 \, \mu\Eh$ is estimated for the $N_\text{b}=1500$ PEC near the $2 \, a_0$ internuclear distance.

Approaching the bond dissociation limit, at $\rho=100 \, a_0$, the energy can be compared with the known sum of atomic energies (see Table \ref{atomvals}):
\begin{align}
 U_\text{a}(\rho=100 \, a_0)&=-5.078 \, 953 \, 551\ ,  \\ 
 U(1 \, ^1S)+U(2 \, ^3S)&=-5.078 \, 953 \, 755 \ , 
\end{align}
suggesting a convergence error of $\sim0.2 \, \mu\Eh$ for large $\rho$.

The error of the PEC cancels to a large degree in the lower vibrational excitation intervals, and especially in the rotational intervals and fine-structure splittings. It does affect, however, the accuracy of the ionization energies. 

Table \ref{tab:BOconv} showcases the convergence of the electronic energy of the \atSup\ state at $\rho=2 \, a_0$ as the $N_\text{b}$ number of basis functions is increased, as well as comparisons with results of Full-Configuration Interaction (FCI) computations performed by us (using the Molpro package~\cite{molpro1,molpro2}) and with available data from the literature. 
It is clearly seen that FCI using uncorrelated one-particle Gaussian basis sets struggles to reduce the convergence error below $1 \, \text{m}\Eh$; even with large doubly and triply augmented correlation-consistent basis sets available in the literature, the electronic energy convergence error cannot be reduced below $0.7 \, \text{m}\Eh$, which is not sufficient for precision spectroscopic purposes. Former MRCI computations from the literature similarly lag behind.

\begin{table}[h!]
  \caption{%
    \atSup\ He$_2$ ($\rho=2\, a_0$):
    convergence of the BO energy, in $\Eh$,
    as a function of the $\Nb$ number of symmetry-adapted floating ECG functions, optimized in a variational procedure. Comparison with literature data from Refs.~\cite{PaCaBuAd08,Ya89,ChJeYaLe89} and Full-Configuration Interaction (FCI) computations carried out with Molpro \cite{molpro1,molpro2} are also shown.
    \label{tab:BOconv}
  }
  \begin{tabular}{@{}ll@{}}
    \hline\hline\\[-0.35cm]
      Basis & 
      $U \ [\Eh]$  \\
  \hline\\[-0.35cm]
  \multicolumn{2}{l}{Literature values} \\       
  400 ECG$^\text{a}$ \cite{PaCaBuAd08} & $-$5.150 439 42 \\
  MRCI/basis$^\text{b}$ \cite{Ya89} & $-$5.146 983         \\
  MRCI/basis~\cite{ChJeYaLe89} & $-$5.147 189       \\
  MRCI~\cite{BjMiPaRo98} & $-$5.150 264    \\     
  \hline \\[-0.35cm]
  \multicolumn{2}{l}{Full-CI (Molpro) [this work]} \\    
  basis~\cite{ChJeYaLe89,Ya89} & $-$5.147         \\[0.15cm]
  aug-cc-pVDZ  & $-$5.10            \\
  aug-cc-pVTZ  & $-$5.138           \\
  aug-cc-pVQZ  & $-$5.144           \\
  aug-cc-pV5Z  & $-$5.146           \\[0.15cm]
  daug-cc-pVDZ & $-$5.125           \\
  daug-cc-pVTZ & $-$5.145 4         \\
  daug-cc-pVQZ & $-$5.149 1         \\
  daug-cc-pV5Z & $-$5.150 22        \\[0.15cm]
  taug-cc-pVDZ & $-$5.126           \\
  taug-cc-pVTZ & $-$5.145 7         \\
  taug-cc-pVQZ & $-$5.149 3         \\
  taug-cc-pV5Z & $-$5.150 36        \\
  \hline\\[-0.35cm]
  \multicolumn{2}{c}{Variational fECG (QUANTEN) [this work]$^\text{c}$} \\        
   10  & $-$5.11            \\
   20  & $-$5.131            \\
   50  & $-$5.147            \\
  100  & $-$5.150 0          \\
  200  & $-$5.150 8          \\
  500  & $-$5.151 07         \\
 1000  & $-$5.151 11         \\
 1500$^\text{d}$  & $-$5.151 122 34 \\
 1500  & $-$5.151 122 50     \\
 2000  & $-$5.151 124 02   \\
 2500 &  $-$5.151 124 69 \\
    \hline\hline\\[-0.35cm]
  \end{tabular}
  ~\\
  $^\text{a}$ This computation corresponds to $R=1.976 \, a_0$. \\
  $^\text{b}$ A PEC is available at this level of theory in the literature.  \\
  $^\text{c}$ The number of symmetry-adapted fECG functions ($\Nb$) is listed in the first column. \\
  $^\text{d}$ This basis set was used for the \atSup\ PEC generation in this work. \\
\end{table}

\begin{table}[h!]
    \caption{Atomic (clamped-nucleus) energies and some expectation values for the $1 \, ^1S$ and $2 \, ^3S$ states of He; all values in atomic units. Mind the factor of $Z=2$ in the definition of $\hat{\delta}_1$. \label{atomvals}}
    \centering
    \begin{tabular}{@{}cr|d{4.10}d{4.10}@{}}
\hline\hline \\[-0.35cm]    
  &  & \multicolumn{1}{c}{$1 \, ^1S$} & \multicolumn{1}{c}{$2 \, ^3S$} \\
\hline \\[-0.35cm]
     $U$ & \cite{AzBeKo18} & -2.903 724 377    & -2.175 229 378 \\
     $\langle r_1^2+r_2^2\rangle$ & \cite{Drake_AMO,BrScHe93} & 2.386 966 0   & 22.928 759 5 \\
     $\langle\hat{\boldsymbol{P}}_\text{el}^2\rangle$ & \cite{Dr88} & 6.12558770 & 4.36534302 \\
    $\langle \hat{H}_\text{MV}\rangle$ & \cite{Drake_AMO,Dr88} & -13.522 016 81 & -10.458 885 19 \\
    $\langle \hat{\delta}_1\rangle$ & \cite{Drake_AMO,Dr88} & 7.241 717 27 & 5.281 420 33 \\
    $\langle \hat{\delta}_2\rangle$ & \cite{Drake_AMO,Dr88} & 0.106 345 37 & 0.000 000 00 \\
    $\langle \hat{H}_\text{OO}\rangle$ & \cite{Drake_AMO,Dr88} & -0.139 094 69 & -0.001 628 43 \\
   $\langle {\cal{P}}(r_{12}^{-3})\rangle$ & \cite{Dr88}  & 0.989 272 45 & 0.038 861 49 \\
    $\ln(k_0)$ & \cite{Ko19} & 4.37016022 & 4.36403682 \\
\hline\hline \\    
    \end{tabular}
\end{table}

\subsection{Relativistic, QED and nonadiabatic corrections}

Two techniques were used to accelerate the (otherwise very slow) basis set convergence of singular operators $\hat{\delta}_1$, $\hat{\delta}_2$, $\hat{H}_\text{MV}$ and ${\cal{P}}(1/r^3)$.

The so-called `Drachmanization' is the rewriting of expectation values of singular operators via identities which would hold for the exact non-relativistic wave function \cite{Dr81,PaCeKo05}. The result is an expectation value with an operator much less localized around electron-nucleus or electron-electron coalescence points, leading to highly improved basis set convergence. We used a numerical Drachmanization technique (numDr)~\cite{RaFeMaMa24,paper-he2p} (a numerically robust adaptation of Drachmanization, enabling expectation value computations for polyatomic systems) to compute the expectation values of all the above singular operators along the PEC.

In the Integral Transformation (IT) approach \cite{PaCeKo05,JeIrFeMa22}, the expectation value is recast as an integral which is split into low-range and high-range regions by a cutoff parameter $\Lambda$; the low-range part is evaluated numerically, while the high-range part is obtained by fitting a semi-analytical asymptotic expansion (the relevant formulae can be derived along the lines of  Ref. \cite{JeIrFeMa22}; mind the factor of $2$ misprint in Eq.~(19) of Ref.~\cite{PaCeKo05}, later corrected in Ref.~\cite{CePrKoMeJeSz12}). 

Table~\ref{tab:relHe2a} shows the convergence of the relativistic corrections, the spin-independent and spin-dependent Breit-Pauli expectation values and the Araki-Sucher term at $\rho=2 \, a_0$.
While the convergence test was carried out for both convergence acceleration techniques, we preferred numDr to compute the corrections along the complete PEC, the main reason being the high sensitivity of IT to the choice of cutoff and fitting parameters. At the same time, numDr requires no manual adjustments, making it more ideal for PEC computations. At $\rho=2 \, a_0$, we estimate the relativistic correction (numDr value) to be accurate to 0.007~$\cm$; the relative error it causes in the rotational and rovibrational intervals is thought to be very small. The main cause of the error in the relativistic corrections is $\langle\hat{H}_\text{MV}\rangle$ which $-$ even in its Drachmanized form $-$ contains a term sensitive to the electron-electron cusp condition (see \emph{e.g.,} Ref.~\cite{PuKoPa17}). 
\newline
At very large internuclear distances, the computed expectation values were found to be in good agreement with the sums of atomic values displayed in Table \ref{atomvals} (the largest, $\sim3\cdot10^{-4} \, \text{a.u.}$ discrepancy shown by $\langle\hat{H}_\text{MV}\rangle$ due to the aforementioned cusp problem, the others accurate to $\sim6\cdot10^{-5} \, \text{a.u.}$). For $\rho>10 \, a_0$, the numerical values of the Araki-Sucher term were compared to the asymptotic formula \cite{Pa05lr}
\begin{equation}
 \left\langle{\cal{P}}(r^{-3})\right\rangle=\left\langle{\cal{P}}(r_{12}^{-3})\right\rangle_{1 \, ^1S}+\left\langle{\cal{P}}(r_{12}^{-3})\right\rangle_{2 \, ^3S}+\frac{4}{\rho^3}+2\frac{\left\langle r_1^2+r_2^2\right\rangle_{1 \, ^1S}+\left\langle r_1^2+r_2^2\right\rangle_{2 \, ^3S}}{\rho^5}+{\cal{O}}(\rho^{-6}) \ ;
\end{equation}
see Table \ref{atomvals} for the above atomic values. The other three singular expectation values decay to their dissociation limits as $\sim\text{const}\cdot\rho^{-6}$.

Computation of the Bethe logarithm is similarly plagued by very slow basis set convergence. The approach of Schwartz~\cite{Sc61,KoHiKa13,Ko19,FeMa22bethe} circumvents this problem by reexpressing $\ln(k_0)$ as a principal value integral over photon momenta, where the integrand can be accurately represented in a suitably optimized auxiliary basis set. The downside of this method is that auxiliary basis sets should be optimized for several values of photon momenta at each point along the PEC, making the computation very time consuming. Numerical observations about the weak dependence of $\ln(k_0)$ on ${\nel}$ and the electronic quantum numbers motivated an ion-core approximation in Ref.~\cite{FeKoMa20}, where the Bethe logarithm of $\text{He}_2^+(^2\Sigma_\text{u}^+)$ was approximated with that of $\text{He}_2^{3+}(^2\Sigma_\text{g}^+)$ (a one-electron state). To avoid the costly evaluation of $\ln(k_0)$ using the Schwartz method, we use the same approximation for the \atSup\ state as well. Some reference values for atomic Bethe logarithms, as well as the ion-core and the rigorously calculated values for the \atSup\ diatomic state (for $\rho=2~a_0$) are displayed in the Table \ref{tab:Bethelog}: a comparison of the values suggests the error of the ion-core approximation to be at most $0.025 \, \text{a.u.}$, which is more than sufficient for the present computations. The effect of the approximate Bethe logarithm on the rovibrational intervals was checked by recomputing the intervals with  $\ln(k_0)$ values slightly perturbed according to the above estimated error. The variance of vibrational and rotational intervals turned out to be only $\sim 5\cdot10^{-4} \, \text{cm}^{-1}$ and $\sim 1\cdot10^{-4} \, \text{cm}^{-1}$ respectively, which are well below the total estimated errors (see Tables II and III of the paper).

Nonadiabatic mass corrections involve the inverse of $\hat{H}_\el-U_\text{a}$ reduced to the electronic subspace orthogonal to \atSup. Computing this reduced resolvent as a sum-over-states expression in the direct basis set approach is not sufficient. Similarly to the previously discussed case of the Bethe logarithm, the computation of mass corrections is recast as an optimization problem \cite{MaFe22nad}. An auxiliary basis set of $^3\Pi_\text{u}$ symmetry was optimized for each $\rho$ to accurately represent the reduced resolvent, leading to results far superior to the direct expansion ones. 

For the sake of completeness, the basis set convergence of $\langle\hat{L}_x^2\rangle=\langle\hat{L}_y^2\rangle$ (needed to evaluate the angular momentum part of DBOC) is showcased in Table \ref{tab:L2conv}.

The computational effort to obtain the relativistic, QED and diagonal BO corrections was negligible compared to the effort of developing the PEC. The non-adiabatic masses required further, optimization of auxiliary basis sets at each point, but were computed at fewer points than the PEC.

\begin{table*}[h!]
  \caption{%
    Convergence of the expectation values of singular operators, 
    $\hat{H}_\text{MV}$, $\hat{\delta}_1$, $\hat{\delta}_2$, and ${\cal{P}}(1/r^3)$ term at $\rho=2 \, a_0$, computed with the electronic wave functions, with the IT technique \cite{PaCeKo05,JeIrFeMa22}, and with the numerical Drachmanization approach \cite{RaFeMaMa24}. The expectation value of $\hH_\text{SS,dp}$ is also shown. All quantities in atomic units; mind that the factor of $\alpha^2$ is not part of the definition of the operators. 
    \label{tab:relHe2a}
  }
    \centering
    \begin{tabular}{@{}r@{\ \ \ \ }c@{\ \ \ \ }c@{\ \ \ \ }c@{\ \ \ \ } c@{\ \ \ \ }c@{\ \ \ \ }c@{\ \ \ \ }c@{\ \ \ \ }c@{}}
    \hline\hline\\[-0.35cm]
       $\Nb$ & 
       $\langle \hat{H}_{\rm MV} \rangle$ &
       $\langle \hat{\delta}_1 \rangle$ &
       $\langle \hat{\delta}_2 \rangle$ &       
       $\langle \hH_\OO \rangle$ $^\text{a}$ &   
       $\langle\hat{H}_\text{BP,ns}\rangle$ & $\langle{\cal{P}}(1/r^3)\rangle$ &
       $\langle\hH_\text{SS,dp}\rangle_{\Sigma=0}$  \\
       \hline\\[-0.35cm] 
       \multicolumn{8}{l}{Direct evaluation} \\
  10      & $-$20.470 259 & 10.594 272 & 0.154 908 & $-$0.068 891 & $-$3.411 047 & $-$ & 0.002 662 0 \\
  20      & $-$21.672 090 & 11.236 577 & 0.158 337 & $-$0.067 345 & $-$3.591 632 & $-$ & 0.003 194 7 \\
  50      & $-$22.918 119 & 11.923 392 & 0.133 784 & $-$0.076 626 & $-$3.845 230 & $-$ & 0.003 407 5 \\
 100      & $-$23.446 039 & 12.230 364 & 0.126 681 &	$-$0.074 448 & $-$3.911 096 & $-$ & 0.002 044 3 \\
 200      & $-$23.754 668 & 12.413 650 & 0.124 398 &	$-$0.073 750 & $-$3.938 295 & $-$ & 0.004 033 1 \\
 500      & $-$23.974 505 & 12.549 806 & 0.121 691 & $-$0.073 281 & $-$3.952 295 & $-$ & 0.004 138 8 \\
1000      &	$-$24.125 596 &	12.643 895 & 0.120 488 & $-$0.073 179 & $-$3.959 267 & $-$ & 0.004 157 9 \\
1500$^\text{b}$ &	$-$24.139 464 &	12.644 903 & 0.120 383 & $-$0.073 154 & $-$3.971 859 & $-$ & 0.004 163 1 \\
1500      &	$-$24.127 286 &	12.644 822 & 0.120 380 & $-$0.073 154 & $-$3.959 815 & $-$ & 0.004 163 1 \\
2000      &	$-$24.128 742 &	12.644 728 & 0.120 369 & $-$0.073 149 & $-$3.961 448 & $-$ & 0.004 164 1 \\
       \hline\\[-0.35cm]
       \multicolumn{8}{l}{IT regularization$^\text{c}$} \\       
             10 & $-$21.343 095 & 10.940 445 & 0.153 714 & $-$ & $-$3.743 868 & 1.012 551 & $-$ \\
             20 & $-$22.606 933 & 11.607 445 & 0.157 116 & $-$ & $-$3.947 752 & 0.980 614 & $-$ \\
             50 & $-$23.882 968 & 12.302 859 & 0.132 752 & $-$ & $-$4.217 254 & 1.208 056 & $-$ \\
            100 & $-$24.017 222 & 12.540 660 & 0.125 701 & $-$ & $-$3.997 962 & 1.357 992 & $-$ \\
            200 & $-$24.127 400 & 12.624 552 & 0.123 423 & $-$ & $-$3.982 805 & 1.415 585 & $-$ \\
            500 & $-$24.148 988 & 12.654 083 & 0.120 777 & $-$ & $-$3.965 851 & 1.509 505 & $-$ \\
           1000 & $-$24.156 022 & 12.660 491 & 0.120 131 & $-$ & $-$3.964 746 & 1.544 389 & $-$ \\
1500$^\text{b}$ & $-$24.156 655 & 12.660 736 & 0.120 061 & $-$ & $-$3.965 189 & 1.546 684 & $-$ \\
           1500 & $-$24.156 641 & 12.660 742 & 0.120 058 & $-$ & $-$3.965 174 & 1.546 787 & $-$ \\
           2000 & $-$24.156 776 & 12.660 883 & 0.120 039 & $-$ & $-$3.965 145 & 1.547 470 & $-$ \\
        \hline\\[-0.35cm]
       \multicolumn{8}{l}{Numerical Drachman regularization$^\text{d}$} \\ 
  10      & $-$23.263 942 & 12.279 549 & 0.108 297 & $-$ & $-$3.703 936 & 1.690 092 & $-$ \\
  20      & $-$23.772 303 & 12.477 963 & 0.110 281 & $-$ & $-$3.892 853 & 1.687 252 & $-$ \\
  50      & $-$24.042 245 & 12.609 270 & 0.117 824 & $-$ & $-$3.942 120 & 1.587 434 & $-$ \\
 100      & $-$24.115 079 & 12.641 376 & 0.119 263 & $-$ & $-$3.957 825 & 1.564 633 & $-$ \\
 200      & $-$24.144 809 & 12.653 864 & 0.119 691 & $-$ & $-$3.965 894 & 1.558 110 & $-$ \\
 500      & $-$24.155 280 & 12.659 452 & 0.119 914 & $-$ & $-$3.966 418 & 1.552 950 & $-$ \\
1000      & $-$24.157 288 &	12.660 973 & 0.119 957 & $-$ & $-$3.965 802 & 1.551 744 & $-$ \\
1500$^\text{b}$ & $-$24.157 271 &	12.660 991 & 0.119 967 & $-$ & $-$3.965 700 & 1.551 582 & $-$ \\
1500      & $-$24.157 273 &	12.660 991 & 0.119 967 & $-$ & $-$3.965 700 & 1.551 577 & $-$ \\
2000      & $-$24.157 350 &	12.661 007 & 0.119 969 & $-$ & $-$3.965 742 & 1.551 544 & $-$ \\
    \hline\hline
    \end{tabular}
    \begin{flushleft}
      $^\text{a}$ From direct computation, no regularization is needed. \\
  $^\text{b}$ This basis set was used for the \atSup\ PEC generation in this work. \\
      $^\text{c}$ See Refs.~\cite{PaCeKo05,JeIrFeMa22} for further numerical details. \\
      $^\text{d}$ See  Ref.~\cite{RaFeMaMa24} for further details.

    \end{flushleft}
\end{table*}

\begin{table*}[h!]
  \caption{%
    Some reference Bethe logarithm values for He$^+$, He, He$_2$ \atSup{}, 
    He$_2^{+}$ X~$^2\Sigma_\text{u}^+$, 
    and 
    He$_2^{3+}$ X~$^2\Sigma_\text{g}^+$. Comparing the atomic and molecular ion values with the rigorous single-point computation of ln$(k_0)$ carried out for \atSup\ in this work demonstrates the insensitivity of the Bethe logarithm to the number of electrons and the appropriateness of the ion-core approximation used in this work.
    \label{tab:Bethelog}
  }
    \centering
    \begin{tabular}{@{}lc clr@{}}
    \hline\hline\\[-0.35cm]
      System      & State    & $R$ [$a_0$]  & $\ln(k_0)$  & \\
      \hline\\[-0.35cm]
      H           & $1\ ^2$S & --    & 2.984 128 555 765     & \cite{DrSw90} \\
      He$^+$      & $1\ ^2$S & --    & 4.370 422 916 885 = $\ln(k_0)_\text{H}+2\ln(2)$     &  \\
      He          & $1\ ^1$S & --    & 4.370 160 223 0703(3) & \cite{Ko19} \\
      He          & $2\ ^1$S & --    & 4.366 412 726 417(1)  & \cite{Ko19} \\  
      He          & $2\ ^3$S & --    & 4.364 036 820 476(1)  & \cite{Ko19} \\
      He          & $2\ ^3$P$_\text{o}$ & --    & 4.369 985 364 549(3)  & \cite{Ko19} \\
      He$_2$\     & \atSup\  & 2.000 & 4.364(5)      & this work  \\
      He$_2^{+}$  & \XdSgp\  & 2.000 & 4.373(1) & \cite{FeMa22bethe} \\
      He$_2^{3+}$ & \XdSgp\  & 7.000 & 4.370 440 87 & \cite{FeKoMa20} \\
      He$_2^{3+}$ & \XdSgp\  & 2.000 & 4.388 148 18 & \cite{FeKoMa20} \\
    \hline\hline
    \end{tabular}
\end{table*}

\begin{table}[h!]
  \caption{%
    Convergence of the expectation value of $\hat{L}_x^2$ or $\hat{L}_y^2$ for the He$_2$ \atSup{} state at $\rho=2 \, a_0$.
  }
    \label{tab:L2conv}
    \centering
    \begin{tabular}{r@{\ \ \ \ }c}
    \hline\hline\\[-0.35cm]
    $N_\text{b}$ & $\langle\hat{L}_x^2\rangle=\langle\hat{L}_y^2\rangle$ \\\hline\\[-0.35cm]
   10      & 3.250 441 \\
   20      & 3.259 685 \\
   50      & 3.252 560 \\
  100      & 3.254 401 \\
  200      & 3.255 369 \\
  500      & 3.256 222 \\
 1000      & 3.256 547 \\
 1500(PEC) & 3.256 570 \\
 1500      & 3.256 573 \\
 2000      & 3.256 582 \\
     \hline\hline   
    \end{tabular}
\end{table}


\end{document}